\documentclass[epsfig,11pt]{article}

\usepackage{epsfig}
\usepackage{graphicx}
\usepackage{array}
\usepackage{caption}
\usepackage{subcaption}
\usepackage{slashed}
\usepackage{amsmath}
\usepackage{braket}
\usepackage{comment}
\usepackage{verbatim}
\usepackage{amssymb}

\newcommand{\beq}{\begin{equation}}   
\newcommand{\eeq}{\end{equation}}
\newcommand{\beqn}{\begin{eqnarray}}   
\newcommand{\eeqn}{\end{eqnarray}}

\newcommand{\bea}{\begin{eqnarray}}
\newcommand{\eea}{\end{eqnarray}}
\newcommand{\be}{\begin{equation}}
\newcommand{\ee}{\end{equation}}
\newcommand{\bead}{\begin{aligned}}
\newcommand{\eead}{\end{aligned}}

\newcommand{\nn}{\nonumber}

\newcommand{\gsim}{\lower.7ex\hbox{$
\;\stackrel{\textstyle>}{\sim}\;$}}
\newcommand{\lsim}{\lower.7ex\hbox{$
\;\stackrel{\textstyle<}{\sim}\;$}}
\begin{document}

\begin{titlepage}
\leftline{UMN--TH--3352/14}
\leftline{FTPI--MINN--14/27}
\vspace{0.7cm}

\begin{center}
{  \large \bf  Low energy dynamics of $U(1)$ vortices in systems with cholesteric vacuum structure}
\end{center}
\vspace{0.6cm}

\begin{center}
 {\large 
Adam Peterson,$^a$ Mikhail Shifman,$^b$ and Gianni Tallarita$^c$}
\end {center}

\vspace{3mm}
 
\begin{center}

$^a${\em University of Minnesota, School of Physics and Astronomy,
Minneapolis, MN 55455, USA}\\[1mm]
$^{b}${\em William I. Fine Theoretical Physics Institute, University of Minnesota,
Minneapolis, MN 55455, USA}\\[1mm]
$^c${\em Centro de Estudios Cient\'{i}ficos (CECs), Casilla 1469, Valdivia, Chile}
\\[1mm]

\end {center}

\vspace{2cm}

\begin{center}
{\large\bf Abstract}
\end{center}

We discuss flux tubes in systems with $U(1)$ gauge, and spin-orbit locked $SO(3)_{S+L}$ symmetry.  The spin-orbit locking is achieved explicitly in the Lagrangian by introducing a parity violating twist term which causes the spontaneous breaking of $SO(3)_{S+L} \rightarrow SO(2)$.  Additionally, this term causes a spontaneous breaking of the translational symmetry along a particular direction.  Thus, the system appears with a cholesteric vacuum under certain conditions of the parameter space.  With this term, the system admits $U(1)$ topologically stable vortices with additional structure in the vortex cores.  This added structure leads to additional moduli appearing in the low energy dynamics.  We determine these solutions and their low energy theory.

\hspace{0.3cm}

\end{titlepage}
%%%%%%%%%%%%

%%%%%%%%%%%%%%%%%%%%%%%%%%%%%%%%
%%%%%%%new definitions %%%%%%%%%%%%%%%%
%%%%%%%%%%%%%%%%%%%%%%%%%%%%%%%%

\section{Introduction}
\label{intro}

Both, superconductivity of the second kind, with the Abrikosov flux tubes inherent to it, and cholesteric crystals (also known as chiral nematic crystals) are among the most venerable physical discoveries which acquired multiple practical applications. Recent studies of the so-called non-Abelian flux tubes (which originally emerged in supersymmetric field theories, see e.g. \cite{Shifman:2009} for a review) led us to consider models which can have both phenomena simultaneously. The ground state in such models presents a superconductor with a cholesteric structure.  Topological defects of the flux tube type that are supported in these models carry (non-Abelian) moduli, i.e. gapless (or quasigapless) excitations of rotational type. A related dynamical pattern that arises in condensed matter physics was discussed
in \cite{Carlson:2003}.

The specific model we consider is characterized by the symmetry group 
\be
G = U(1)_{\rm gauge} \times SO(3)_{S+L} \times T,
\label{SymmetryGroup}
\ee
where $SO(3)_{S+L}$ is the spin-orbit locked symmetry group of rotations, and $T$ is the translational group in three spatial dimensions.  The spin-orbit locking is achieved with an explicit parity violating term added to the Lagrangian, that preserves only the locked subgroup $SO(3)_{S+L} \subset SO(3)_S \times SO(3)_L$.  An additional feature of this term is the spontaneous breaking of the spatial translational symmetry along a particular direction when certain conditions of the parameters are met.  We choose this direction as our $z$ axis:
\be
T \rightarrow T_x \times T_y.
\ee
Additionally, this term causes the spontaneous breaking of $SO(3)_{S+L}$.  However, a particular combination of the $z$ translations and the $SO(2)$ rotations about the $z$ axis will be preserved by the vacuum.

The spontaneous breaking of the $U(1)$ gauge symmetry leading to the Abrikosov-Nielsen-Olsen flux-tubes in the vacuum is well reviewed \cite{Abrikosov:1957}.  Our goal in this paper is to discuss the response of the additional fields appearing in the model when the added parity violating term is included.  As this additional term does not involve either the Higgs or the gauge vector field, the $U(1)$ charged vortices will remain topologically stable in this model.  They will however acquire additional structure due to the breaking of translational symmetry in the vacuum.  

When the parity violating term is small we may discuss the perturbative effect on the low energy dynamics of the vortex.  It is well known that the dynamics of vortices in the $U(1)$ model without the parity violating term reduce to a $1+1$ dimensional model of excitations propagating along the vortex axis.  These propagating modes follow from the translational symmetry breaking (Kelvin excitations), and other broken non-Abelian rotational symmetries which typically follow the dynamics of a sigma model (see \cite{Shifman:2013a}).  In the case of a broken $SO(3)$ rotational symmetry, the additional excitations are described by a $CP(1)$ model on the vortex.  We will see that the additional parity violating term will lift some of the $CP(1)$ degrees of freedom.

We will begin by discussing the model in general and introducing the parity violating twist term.  The following section will be devoted to exploring the vacuum structure of the model and determining the conditions on the parameter space required for each vacuum to be the minimum.  Section 5 will discuss the vortices and discuss the details of the numerical procedure used to obtain solutions.  Section 6 will develop the low energy dynamics of the vortices.  We will discuss the results and conclude in section 7.

\section{The system}

The simplest model supporting flux tubes of the Abrikosov type, with non-Abelian excitations localized on the tube was suggested in \cite{Shifman:2013} based on an extension of \cite{Witten:1985} and further studied in \cite{Shifman:2013a}. Its Lagrangian has the form
\begin{align}
\mathcal{L}  &=\mathcal{L}_0+\mathcal{L}_\chi \notag \\[2mm]
\mathcal{L}_0 &= -\frac{1}{4e^2}F_{\mu\nu}F^{\mu\nu}+|D_\mu \phi|^2-\lambda(|\phi|^2-v^2)^2 \notag \\[2mm]
\mathcal{L}_\chi &=\partial_t\chi_i\partial_t \chi_i - \partial_i \chi_j \partial_i \chi_j-\gamma \left[(-\mu^2+|\phi|^2)\chi_i\chi_i+\beta(\chi_i\chi_i)^4\rule{0mm}{4mm}\right],
\label{Lagrangian}
\end{align}
where the subscript $i$ runs over $i=1,2,3$. The field $\chi_i$ can be viewed as a spin field. The covariant derivative is defined in a standard way
\be
D_\mu = \partial_\mu- i A_\mu\,.
\ee
Under an appropriate choice of parameters the charged field $\phi$ is condensed in the ground state,
\be
|\phi|_{\rm vac} = v\,.
\ee
Moreover, if $\mu<v$, the field $\chi$ is not excited in the vacuum
\be
\left(\chi_i\right)_{\rm vac} = 0\,.
\ee
The model obviously supports the Abrikosov flux tube. Inside the tube, the spin field $\chi_i$ is excited giving rise to gapless (or quasigapless) excitations of non-Abelian type, localized on the flux tube.

Now we would like to make the next step and introduce a twist term $\mathcal{L}_\varepsilon$ which violates parity through mixing of the ``spin term" with angular momentum, 
\begin{equation}
{\mathcal L} = \mathcal{L}_0+\mathcal{L}_\chi +\mathcal{L}_\varepsilon\,,\qquad
\mathcal{L}_\varepsilon = -\eta \varepsilon_{ijk}\chi_i \partial_j \chi_k\,,
\label{LagrangianEta}
\end{equation}
where $\eta$ is a deformation parameter.  The spatial kinetic terms of $\chi_i$ including the twist $\mathcal{L}_\varepsilon$, is recognized as the Frank-Oseen free energy density of an isotropic chiral nematic liquid crystal \cite{deGennes:1993}.  Note that the twist term is linear in derivatives. If $\eta$ is large enough, a vacuum expectation value of $\chi_i$ develops with a cholesteric structure.

$\mathcal{L}_\varepsilon$ also breaks the orbital rotational part of the Lorentz symmetry  implying a spin-orbit locked symmetry of the full Lagrangian,
\begin{equation}
SO(3)_L \times SO(3)_S \rightarrow SO(3)_{S+L}.
\end{equation}
The energy density derived from the Lagrangian (\ref{LagrangianEta}) is
\bea
E &=& \frac{1}{4e^2}F_{ij}F^{ij}+|D_i\phi|^2+\lambda(|\phi|^2-v^2)^2+\partial_i\chi_j\partial_i\chi_j \nonumber\\[2mm]
&+& \eta \varepsilon_{ijk}\chi_i \partial_j \chi_k+\gamma \left[(-\mu^2+|\phi|^2)\chi_i\chi_i+\beta(\chi_i\chi_i)^2\right].
\label{EnergyDensityEta}
\eea
Our first task is to study the vacuum (ground state) of the dynamical system described by (\ref{LagrangianEta}) or (\ref{EnergyDensityEta}).

\section{Generalities}

Assuming that all couplings of the model at hand are small in what follows we will solve static classical equations of motion (i.e. we will limit ourselves to the quasiclassical approximation). The Lagrangian (\ref{LagrangianEta}) contains a number of constants: $e$, $\lambda$, $\beta$ and $\gamma$ (dimensioneless couplings) and dimensionful parameters $v$, $\mu$ and $\eta$.  The mass of the elementary excitations of the charged field $\phi$  is
\be
m_{\phi}^2 = 4\lambda v^2\,.
\ee
In the next sections we will rescale all quantities to appear below to make them  dimensionless, for instance, distance in the direction of the flux tube axis
\be
\tilde{z} = m_\phi z\,,
\ee
distance in the perpendicular direction
\be
\rho= m_\phi\sqrt{x^2+y^2}\,,
\ee
and so on.  Other dimensionless parameters are
\be
b= \frac{\gamma(c-1)}{4\lambda c}\,,\quad c=\frac{v^2}{\mu^2}\,,\quad  a=\frac{e^2}{2\lambda}\,, \quad \tilde{\eta}=\eta/m_\phi\,.
\ee
The field $\chi_i$ being represented in Cartesian coordinates takes the form
\be
\chi_i = \frac{\mu}{\sqrt{2\beta}}\, \left\{ \rule{0mm}{4mm} \tilde{\chi}_x(x,y,z),\,\, \tilde{\chi}_y(x,y,z),\,\, \tilde{\chi}_z(x,y,z) \right\}.
\ee

The static classical equations of motion are derived by extremization of  energy (\ref{EnergyDensityEta}), in a general coordinate system they read
\bea
\label{eomm}
&& \nabla_i\left(\sqrt{-g}g^{ij}\nabla_j \chi_k\right)-\eta\varepsilon_{kji}\nabla_j\chi_i-\frac{\partial V}{\partial \chi_k}=0\,,\nn\\[3mm]
&& D_i\left(g^{ij}\sqrt{-g}D_j\phi\right)+\sqrt{-g}\left(2\lambda\left(|\phi|^2-v^2\right)+\gamma g^{ij} \chi_i\chi_j\right)\phi=0\,,\nn\\[4mm]
&&\partial_i\left(\sqrt{-g}g^{ji}g^{kl}F_{jk}\right)-ie^2\sqrt{-g}\left(\phi^*D^l\phi-\phi D^l\phi^*\right)=0\,,
\eea
where
\be
\frac{\partial V}{\partial \chi_k} = \sqrt{-g}\gamma\left( \left(-\mu^2+|\phi|^2\right)+2\beta g^{ij}\chi_i\chi_j\right)\chi_k\,,
\ee 
and 
\be
\nabla_i\chi_j = \partial_i\chi_j - \Gamma^k_{ij}\chi_k
\ee
is the standard curved space covariant derivative.

\section{Ground state}
\label{gs}

Inspection of ${\mathcal L}_\varepsilon$ in the Lagrangian (\ref{LagrangianEta}) -- in particular,  the fact that it is of the first order in derivative -- prompts us that, generally speaking, in the ground state translational invariance will be spontaneously broken. One can always assume that this breaking is aligned in the $z$ direction. Then minimization of energy argument suggests that the spin field $\chi_i$ is oriented in the $x,y$ plane and rotates as we move  in the $z$ direction.  In other words, a cholesteric structure appears in the ground state
\bea
\label{CholestericStructure}
\chi_i &=& \chi_0\, \epsilon_i(z)\,,\qquad \, \chi_i =  \frac{\mu}{\sqrt{2\beta}}\tilde\chi_0\, \epsilon_i(z)\,,
\nonumber\\[2mm]
{\vec\epsilon}\,(z) &=& \left\{ \rule{0mm}{4mm} \cos{kz},\,\,\sin{kz},\,\, 0\right\}\,,
\eea
where $k$ is a wave vector (more exactly, the wave vector is $\vec k$, but with our choice of the reference frame it has only the $z$ component, $\vec k\to \left\{ 0,\,\,0,\,\, k\right\}$).
Since the coupling between the fields $\phi$ and $\chi$ occurs only through $\chi_i\chi_i=\chi_0^2$ we can also put
\be
\phi = \phi_0\,, \,\,\, \quad A_i=0\,,
\ee
where $\phi_0 $ is a constant.   Then equations of motion (\ref{eomm}) imply
\bea
&& \chi_0\left(-k^2+\eta k-\gamma((-\mu^2+ \phi_0^2)+2\beta\chi_0^2)\right)=0\,,
\nonumber\\[2mm]
&&\phi_0\left(2\lambda(\phi_0^2-v^2)+\gamma\chi_0^2\right)=0\,.
\label{20}
\eea
First we minimize over the wave vector $k$ and obtain 
\be
k=\eta/2 \label{keta}
\ee
demonstrating that the translational symmetry is unbroken only if $\eta =0$. 

For consistency with later sections of the paper we use here the dimensionless field definitions
\bea
\tilde{E}_{vac} &=& \frac{1}{m_\phi^2v^2}E_{vac}\nn\\
\tilde{\chi}_0 &=& \frac{\sqrt{2\beta}}{\mu}\chi_0\nn\\
\varphi &=& \frac{\phi}{v}.  
\eea

Generally speaking the system of equations (\ref{20}) has three independent solutions representing local minima of the energy,
\begin{itemize}
\item[I)]
\bea
(\varphi^2_0)^{I} = \frac{1-\frac{b}{\beta (c-1)}-\frac{\tilde{\eta}^2}{4 \beta}}{1-\frac{bc}{\beta (c-1)}}\,,\quad (\tilde\chi^2_0)^{I} = (c-1)\frac{\frac{\tilde\eta^2}{4b}-1}{1-\frac{bc}{\beta (c-1)}} \nn,
\label{Vac1}
\eea
\be
\tilde{E}_{vac}^{I} =-\frac{(c-1)^2(\tilde{\eta}^2-4 b)^2}{\beta (c-1) -bc}
\label{E1}
\ee
\item[II)]
\be
\label{Vac2}
(\varphi^2_0)^{II} = 1,\quad \left(\tilde{\chi}_0\right)^{II}=0, \quad \tilde{E}_{vac}^{II} = 0,
\ee
\item[III)]
\be
(\varphi_0)^{III} =0, \quad (\tilde{\chi}^{2}_0)^{III}=1+(c-1)\frac{\tilde{\eta}^2}{4b}, \nn\\
\label{Vac3}
\ee
\be
\tilde{E}_{vac}^{III} = \frac{16 bc(c-1)\beta-((c-1)\tilde{\eta}^2+8b)^2}{64bc(c-1)\beta}.
\label{E3}
\ee
\end{itemize}
In this paper we always assume
\be
\label{ParameterConditions}
c >1 ,\quad b >0, \quad \beta > 0\quad \text{and} \quad a>0.
\ee

It will be necessary to determine which vacuum is the minimum for the given parameters in (\ref{ParameterConditions}).  For the moment we will make the additional assumption that 
\be
\beta(c-1) > bc.
\label{Condition1}
\ee
Vacuum $I$ represents the case where both the translational and $U(1)$ gauge symmetries are broken by the non-zero vacuum expectation values of $\chi_0$ and $\phi_0$.  It is easy to show from (\ref{Vac1}) that this vacuum exists and is a minimum if in addition to (\ref{Condition1}),
\be
4b < \tilde{\eta}^2 < 4\beta\left(1 - \frac{b}{\beta(c-1)}\right).
\label{Condition1Vac1}
\ee

Vacuum $II$ as given in (\ref{Vac2}) implies the breaking of the $U(1)$ gauge symmetry, while preserving the translational symmetry in the vacuum since $\chi_0 = 0$.  When $\tilde{\eta}^2 < \tilde{\eta}_{{\rm crit}_1}^2 \equiv 4b$ vacuum $II$ (\ref{Vac2}) is the energy-minimizing solution.  As $\tilde{\eta}^2$ crosses over $\tilde{\eta}^2_{{\rm crit}_1}$ there is a second order phase transition from vacuum $II$ to vacuum $I$.

We can also see that there is a second critical point 
\be
\tilde{\eta}^2_{{\rm crit}_2} \equiv 4\beta \left(1-\frac{b}{\beta(c-1)}\right),
\ee
where there is another phase transition from vacuum $I$ to vacuum $III$ as $\tilde{\eta}^2$ crosses over $\tilde{\eta}^2_{{\rm crit}_2}$.  As $\phi_0 = 0$ in vacuum $III$ the $U(1)$ gauge symmetry is preserved.  Thus, vortices with $U(1)$ topological charge do not exist in vacuum $III$.  We will briefly discuss this case further in the sections below, however our main focus for the paper will be on vortices in vacua $I$ or $II$.  Figure 1 illustrates the vacuum energy as a function of $\tilde{\eta}$.

We may also consider the condition
\be
bc > \beta(c-1).
\ee
We can see from (\ref{E1}) that the vacuum energy of $I$ is greater than the vacuum energy of $II$ in all cases, and is thus never a minimizing solution.  In this case we can see from (\ref{Vac2}) that the minimizing vacuum is $II$ when
\be
\tilde{\eta}^2 < \tilde{\eta}^2_{{\rm crit}_3}=\frac{\sqrt{(c-1)bc\beta}-b}{(c-1)}.
\ee 
Additionally, when the critical point $\tilde{\eta}^2_{{\rm crit}_3}$ exists, there is a first order transition from vacuum $II$ to vacuum $III$ at that critical point.  If however, 
\be
(c-1)c\beta < b,
\ee
then $\tilde{\eta}^2_{{\rm crit}_3}$ is non-existent, and vacuum $III$ is the only true vacuum for all $\tilde{\eta}^2$.  The vacuum energy for the case $bc > \beta(c-1)$ is shown in Figure 2.  

\begin{figure}[ptb]
\centering
\includegraphics[width=0.8\linewidth]{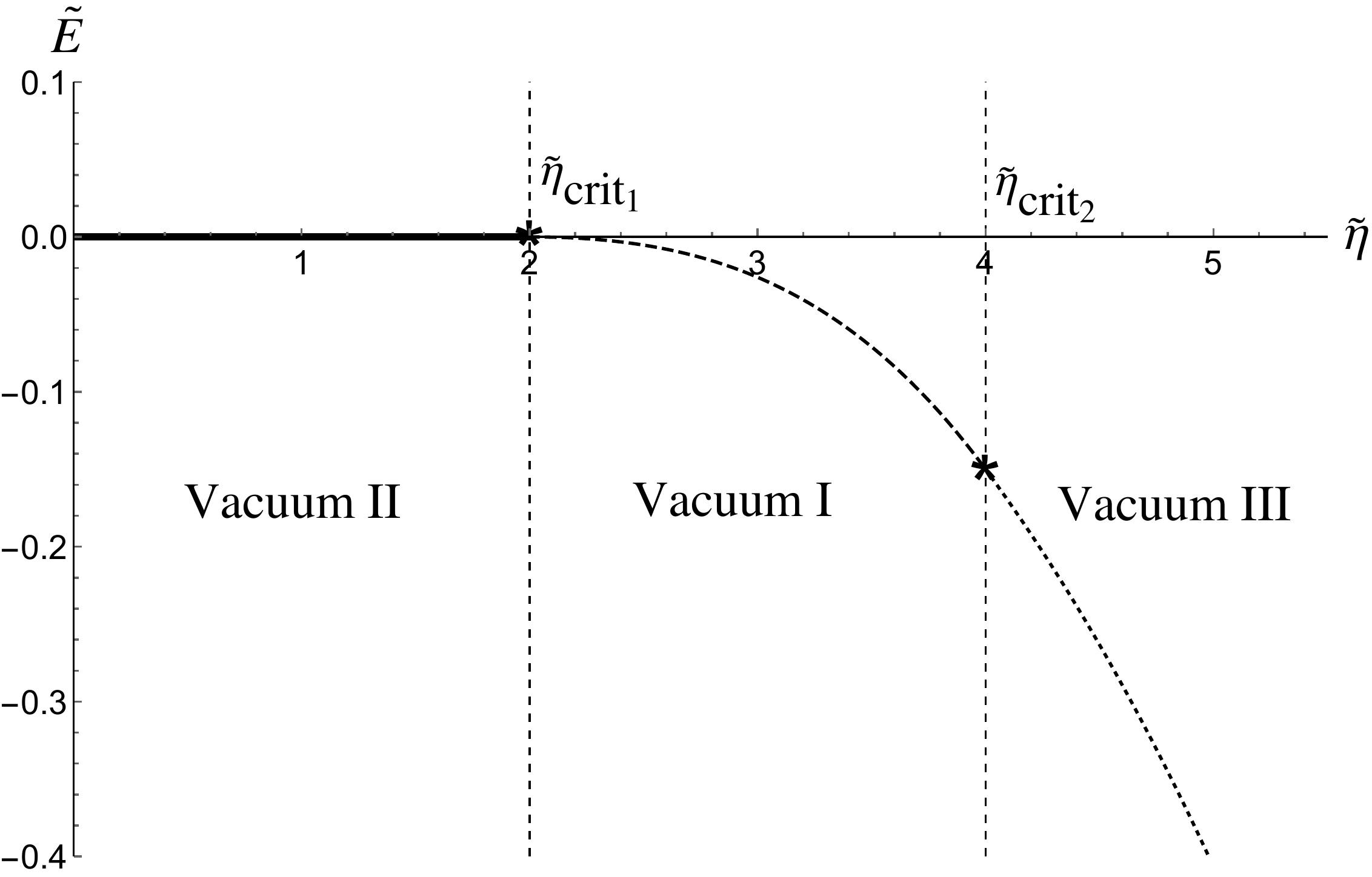}
\caption{$b=1,\; c=1.25,\; \beta=8$. Vacuum energy dependence on $\tilde{\eta}$ for $\beta(c-1)>bc$, the solid line corresponds to vacuum $II$, the dashed line to vacuum $I$ and the dotted line to vacuum $III$.}
\end{figure}

\begin{figure}[ptb]
\centering
\includegraphics[width=0.8\linewidth]{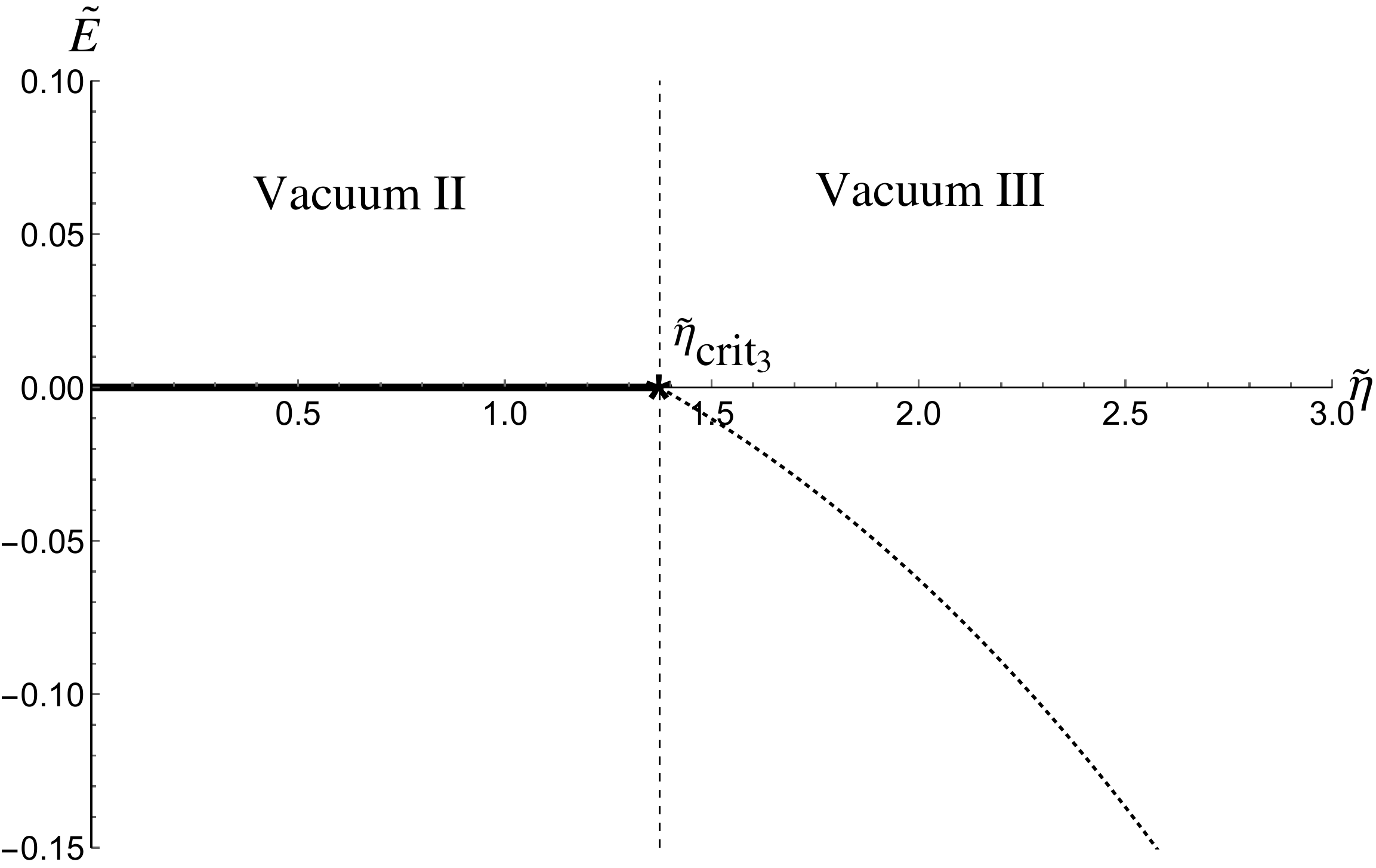}
\caption{$b=1,\; c=1.25,\; \beta=4$. Vacuum energy dependence on $\tilde{\eta}$ for $\beta(c-1)<bc$, the solid line corresponds to vacuum $II$ and the dotted line to vacuum $III$.}%
\end{figure}

Corresponding to the breaking of translational invariance in the vacuum $I$ there exists a Goldstone mode related to spatially dependent phase shifts of the cholesteric structure.  In general, the cholesteric phase breaks all four of the generators of the global translational and rotational generators $p_z$ and $\vec{J}$ while still preserving the linear combination $kJ_z-p_z$.  Thus, we expect a Goldstone mode associated to the broken translational symmetry in the $z$ direction.  Namely if 
\be
\vec{\epsilon}(z) \rightarrow \vec{\epsilon}(x,y,z) = \big\{\cos\left(kz - k\xi(x,y,z)\right),\; \sin\left(kz - k\xi(x,y,z)\right),\; 0\big\},
\label{TranslationalTransformation}
\ee
is inserted into (\ref{Lagrangian}) in which $k=\eta/2$ one finds that 
\be
E = E_{vac} + \frac{\chi_0^2\eta^2}{4} \left(\nabla_i \xi\right)^2
\ee
where $i = (x,y,z)$.  Of course the translation (\ref{TranslationalTransformation}) is equivalent to a rotation of $\epsilon_i$ about the $z$ axis by an angle $\vartheta =k\xi$ , showing the preserved $kJ_z - p_z$ symmetry of the cholesteric vacuum.

Since there are three broken generators by the cholesteric vacuum--$J_x$, $J_y$, and $p_z + kJ_z$--it would be naively expected that three Goldstone modes would appear in the low energy theory instead of one.  However, it can be shown that the orientational moduli from $J_x$ and $J_y$ are lifted from the spectrum for energies much lower than the pitch $\eta/2$ of the vacuum \cite{Radzihovsky:2011}.  Essentially, the additional orientational moduli from $J_{x,y}$ acquire mass gaps due to their mutual interaction and interactions with $p_z+kJ_z$ \cite{Watanabe:2014}.  This effect will also appear on the low energy theory of vortices, which we will describe in section 6 below.

\section{Vortex Solutions}
In this section we will discuss vortices in the three vacua discussed in the previous section.  The model (\ref{LagrangianEta}) admits two types of vortices of which we will discuss only one type in detail.  In vacua I and II, the accumulation of the field $\chi$ in the vortex core can be considered as a response to the effective potential emerging from the topological vortex in $\phi$.  In these cases, the field $\chi$ does not carry a topological charge of its own.  These vortices will be the focus of our attention.  

In contrast, one could also consider a winding of the $SO(3)_{S+L}$ global symmetry by giving $\chi_i$ a $\theta$ dependence at large distances from the vortex axis.  This type of vortex is typically called a spin vortex and arises in models of superfluid $^3$He \cite{Volovik:1977}.  Although, the $SO(3)$ charged vortices in this model present an interesting avenue, we will not discuss them further here.

\subsection{$U(1)$ topological vortices}
In the search of the vortex solution we will switch to polar coordinates.  Note that finding a vortex in the cholesteric vacuum is a nontrivial numerical problem.  We will limit ourselves to the simplest case: the vortex axis parallel to the $z$ axis (or, which is the same, coinciding with $\vec k$). The geometry of our problem is illustrated in Figure 3.

\begin{figure}[ptb]
\centering
\includegraphics[width=0.9\linewidth]{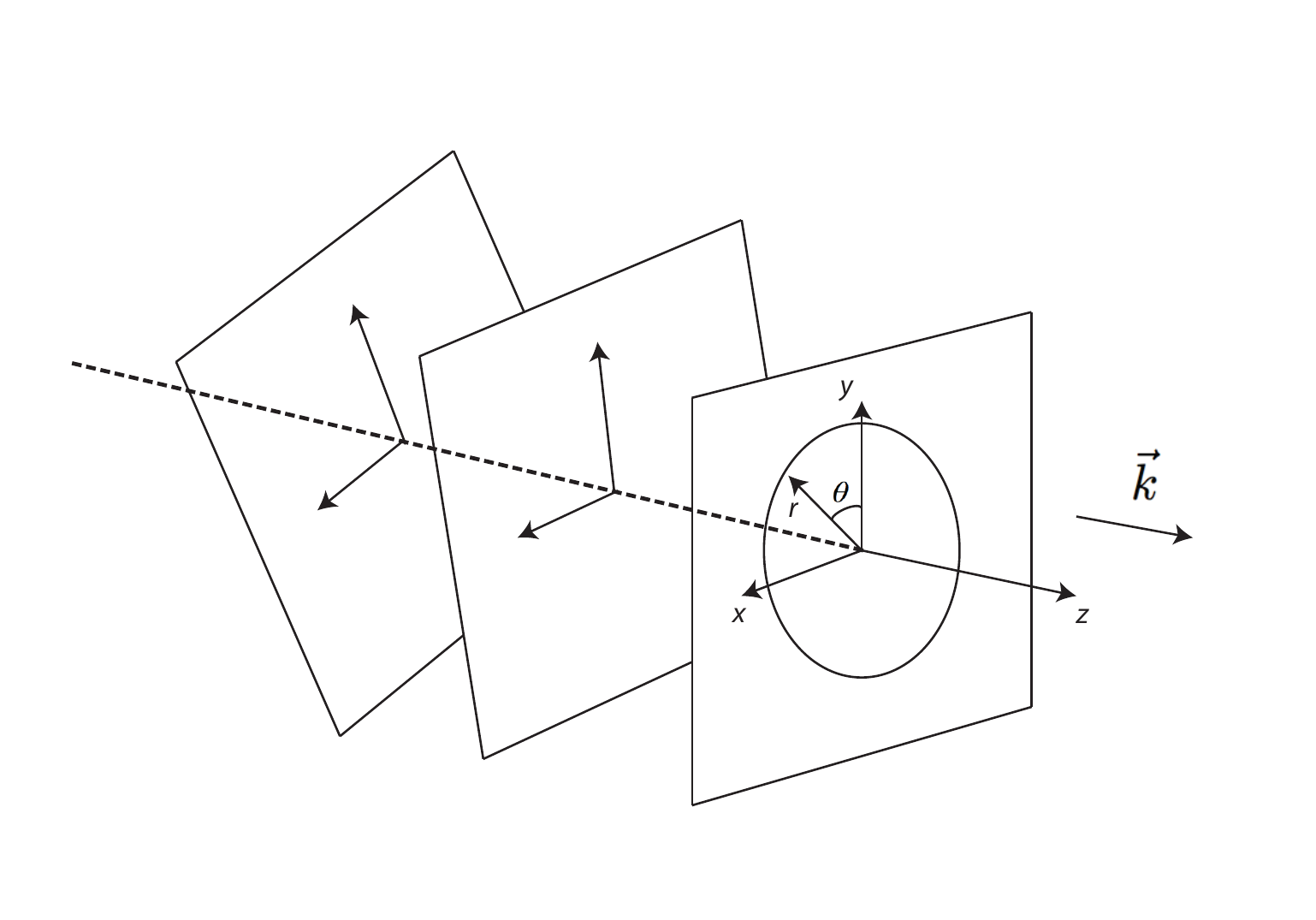}
\caption{Geometric set-up of the problem. The vortex axis, in the $z$ direction, is parallel to the wave-vector $\vec{k}$ which is normal to the cholesteric planes.}%
\end{figure}
We will solve the full three dimensional numerical problem for the functions
\be
\chi_i=\frac{\mu}{\sqrt{2\beta}} \tilde{\chi}_i, \;\;\;\;
\phi = v \varphi(\rho, \theta, \tilde{z}), \;\;\;\; A_i = m_\varphi \tilde{A}_i.
\ee
In the equations to follow, we will omit the tilde above $A$, $\chi$, $z$, and $\eta$ to avoid cluttering the notation.  We will also work in polar coordinates,
\begin{align}
&\vec{\chi}(r,\theta,z) = \chi^r \partial_r+\chi^\theta \partial_\theta +\chi^z \partial_z, \nn\\[2mm]
&\vec{A}(r,\theta,z) = A^r \partial_r+A^\theta \partial_\theta +A^z \partial_z.
\end{align}

Note that $\phi$ (and correspondingly $\varphi$) is a complex function.  Additionally, we work in the gauge $\nabla \cdot \vec{A} = 0$.
Then the equations of motion for $\chi_i$ reduce to
\begin{align}
\frac{1}{\rho}\partial_\rho&\left(\rho\partial_\rho\left(\rho\chi^\theta\right)\right)+\rho\partial^2_{z}\left(\chi^\theta\right)+\frac{1}{\rho}\left(\partial^2_{\theta}\chi^\theta-\chi^\theta+\frac{2}{\rho}\partial_{\theta}\chi^r\right) \nn\\[2mm]
&+\eta\left(\partial_\rho\chi^z-\partial_{z}\chi^r\right)-(V_\chi)\rho\chi^\theta=0,\nn \\[4mm]
\frac{1}{\rho}\partial_\rho&\left(\rho\partial_\rho\chi^r\right)+\partial^2_{z}\chi^r+\frac{1}{\rho^2}\left(\partial^2_{\theta}\chi^r-\chi^r-2\rho\partial_{\theta}\chi^{\theta}\right) \nn \\[2mm]
&+\eta\left(\rho\partial_{z}\chi^\theta -\frac{\partial_\theta\chi^z}{\rho}\right)-(V_\chi)\chi^r=0, \nn \\[4mm]
\frac{1}{\rho}\partial_\rho&\left(\rho\partial_\rho\chi^z\right)+\partial^2_{z}\chi^z+\frac{1}{\rho^2}\partial^2_{\theta}\chi^z-\eta\left(\rho\partial_\rho\chi^\theta+2\chi^\theta-\frac{\partial_{\theta}\chi^r}{\rho}\right) \nn \\[2mm]
&- (V_\chi)\chi^z=0.
\label{ChiSystem}
\end{align}
The equations for the gauge field $A$ reduce to
\begin{align}
\frac{1}{\rho}\partial_\rho&\left(\rho\partial_\rho\left(\rho A^\theta\right)\right)+\rho\partial^2_{z}A^\theta+\frac{1}{\rho}\left(\partial^2_{\theta}A^\theta-A^\theta+\frac{2}{\rho}\partial_{\theta}A^r\right) &\nn\\[2mm]
&-\frac{a\rho}{2}\left(J^\theta+2A^\theta |\varphi|^2\right)=0,\nn\\[4mm]
\frac{1}{\rho}\partial_\rho&\left(\rho\partial_\rho A^r\right)+\partial^2_{z}A^r+\frac{1}{\rho^2}\left(\partial^2_{\theta}A^r-A^r-2\rho\partial_{\theta}A^{\theta}\right)\nn \\[2mm]
&-\frac{a}{2}\left(J^r+2 A^r |\varphi|^2\right)=0,\nn\\[3mm]
\frac{1}{\rho}\partial_\rho&\left(\rho\partial_\rho A^z\right)+\partial^2_{z}A^z+\frac{1}{\rho^2}\partial^2_{\theta}A^z-\frac{a}{2}\left(J^z+2A^z|\varphi|^2\right)=0.
\label{ASystem}
\end{align}
The equation for the scalar field $\varphi$ reduce to
\begin{align}
\frac{1}{\rho}\partial_\rho&\left(\rho\partial_\rho\varphi\right)+\frac{1}{\rho^2}\partial^2_\theta\varphi+\partial^2_{z}\varphi-2\mbox{i}\left(A^r\partial_\rho+A^\theta \partial_\theta +A^z \partial_{z}\right)\varphi \nn\\[2mm]
&-A^2\varphi-(V_\varphi)\varphi=0,
\label{PhiSystem}
\end{align}
where
\begin{align}
& V_\chi = \frac{b}{c-1}\left[-1+c|\varphi|^2+\chi^2\right], \nn\\[4mm]
& V_\varphi = \frac{1}{2}\left(|\varphi|^2-1+\frac{b}{\beta(c-1)}\chi^2\right), \nn\\[4mm]
& J_i = \mbox{i}(\varphi^*\partial_{i}\varphi-\varphi\partial_{i}\varphi^*).
\label{Potentials}
\end{align}

In the polar coordinates, when $\beta(c-1) > 4bc$ and $\eta_{{\rm crit}_1} < \eta < \eta_{{\rm crit}_2}$, the vacuum solution describes a cholesteric vacuum and takes  the form
\be
\chi^r = \chi_0 \cos(kz-\theta),\;\;\; \chi^\theta = \frac{\chi_0}{\rho} \sin(kz-\theta), \;\;\; \chi^z =0,
\ee
where we are assuming $k$ and $z$ are in dimensionless units.  Note that this corresponds to a vacuum which, in terms of Cartesian vectors $\chi_x , \chi_y$ and $\chi_z$, has no $kz-\theta$ dependence. In order to find finite energy solutions in this parameter range we therefore seek solutions to these equations with the following boundary conditions in $\rho$:
\begin{align}
&\chi^z(\rho,\theta, z)\xrightarrow{\rho \rightarrow \infty} 0, \nn\\[2mm]
&\chi^r(\rho, \theta, z)\xrightarrow{\rho \rightarrow \infty} \chi_0 \cos(kz-\theta), \nn\\[2mm]
& \chi^\theta(\rho, \theta, z)  \xrightarrow{\rho \rightarrow \infty} \chi_0 \sin(kz-\theta)/\rho,\nn\\[2mm]
&\varphi(0,\theta,z)=0,\;\; \varphi(\rho,\theta,z) \xrightarrow{\rho \rightarrow \infty} \varphi_0{\rm e}^{{\rm i}\theta},\nn\\[2mm]
&A^\theta(\rho,\theta,z) \xrightarrow{\rho \rightarrow \infty} 0,\nn\\[2mm]
&A^r(0,\theta,z)=0, \;\; A^r(\rho,\theta,z)\xrightarrow{\rho \rightarrow \infty} 0, \nn\\[2mm]
&A^z(\infty,\theta,z)=0.
\label{DynamicalBoundaryConditions}
\end{align}
Additionally, there are boundary conditions on $\chi^r$, $\chi^\theta$, $A^r$ and $A^\theta$, at $\rho \rightarrow 0$ that are required for the solution to be analytic at the origin. The boundary condition on $\varphi(\rho \rightarrow \infty)$ ensures the $U(1)$ winding of the solution.  We also demand periodicity in the angular $\theta$ direction and continuity in the $z$ direction.  Solutions for the field $\chi_i$ with $a=1$, $b=1$, $c=1.25$, $\beta=8$, and $\eta = 2.2$ are shown for $z=0$ in Figures 4 and 5 where the cholesteric structure is clearly visible.  A plot of the density $|\vec{\chi}|$ is shown in Figure 6.  Figure 7 shows a plot of $|\varphi|$ and $A^\theta$ as functions of $\rho$ at $z=\theta=0$ indicating the vortex structure at the origin of the $xy$ plane.  For these values of the parameters $\eta_{{\rm crit}_1}=2$ and $\eta_{{\rm crit}_2}=4$.  In Figures 8, 9, and 10 we show similar plots for the same choice of parameters $a$, $b$, $c$, and $\beta$ but with $\eta = 0.5$.  Thus, these vortices sit in vacuum $II$.  \newline

\begin{figure}
\hspace*{-2.4cm}
\begin{subfigure}{5cm}
\centering
\includegraphics[width=1.5\linewidth]{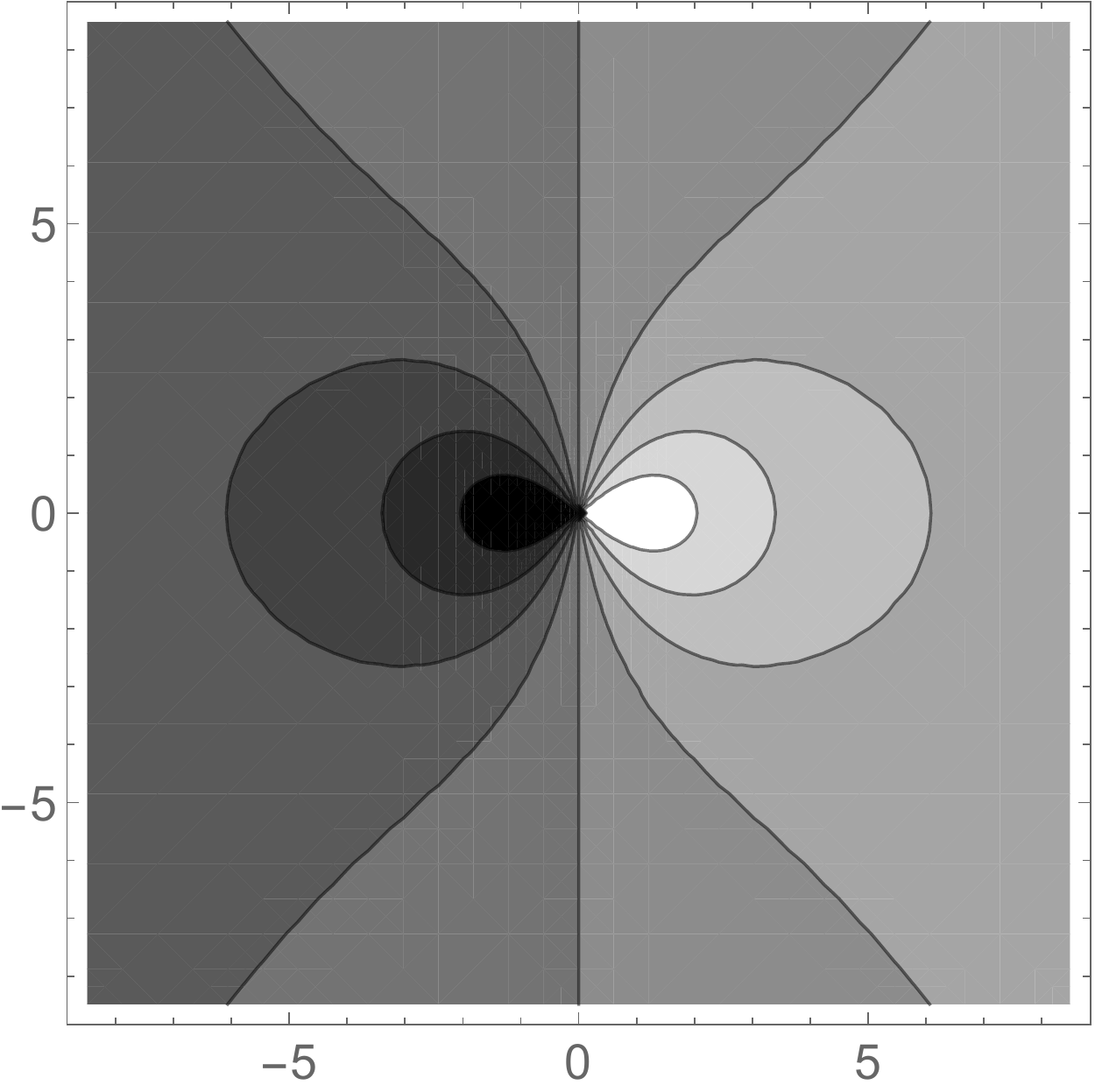}
\caption{$\chi^r$, at $kz=0$.}
\end{subfigure}%
\quad\quad\quad\quad\quad\;\;\;\;\;\;\;\;\;
\begin{subfigure}{5cm}
\centering
\includegraphics[width=1.5\linewidth]{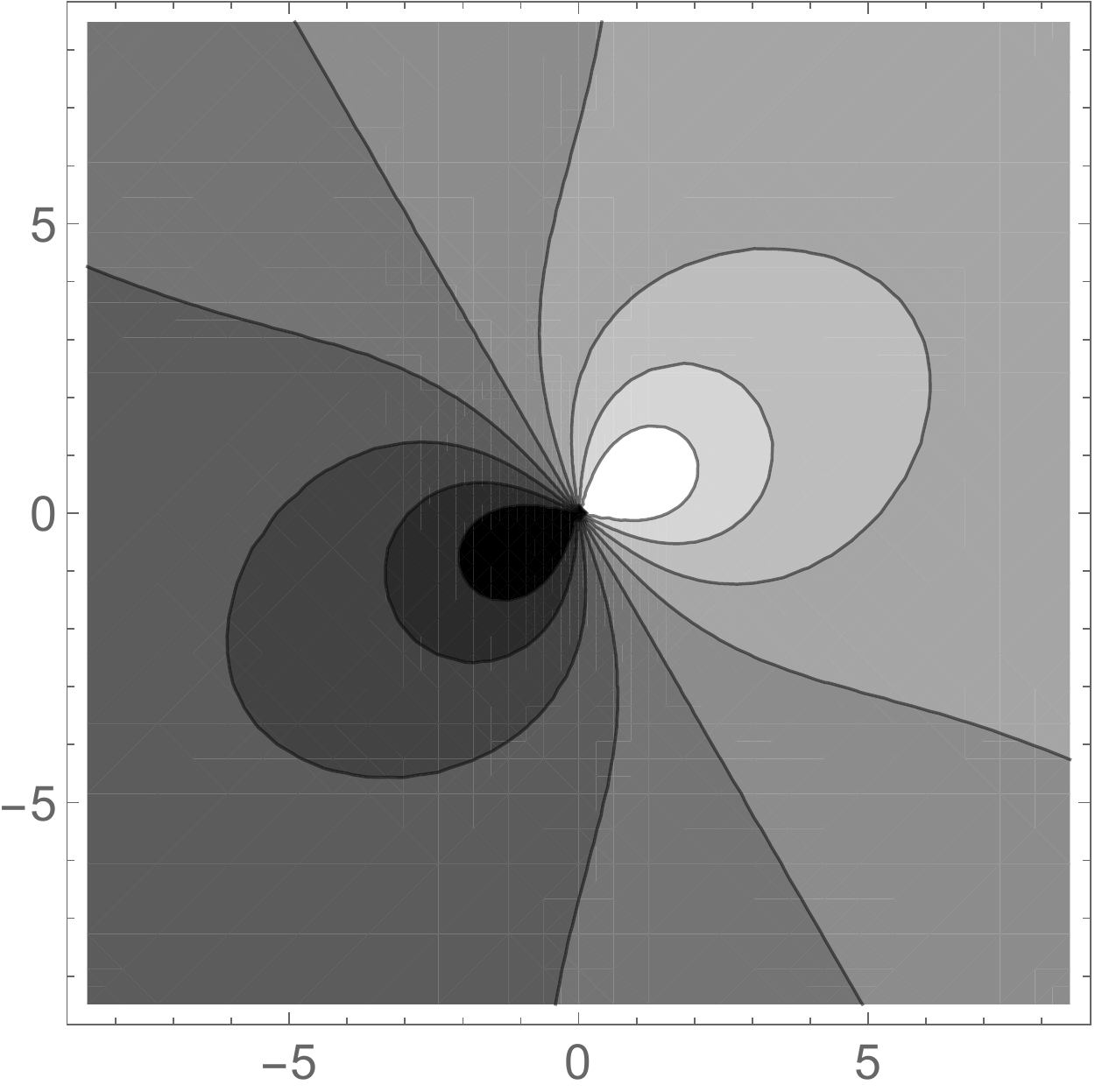}
\centering
\caption{$\chi^r$ at $kz = \pi/6$.}
\end{subfigure}\\
\\
\\
\centering
\hspace*{1.3cm}
\begin{subfigure}{5cm}
\centering
\includegraphics[width=1.5\linewidth]{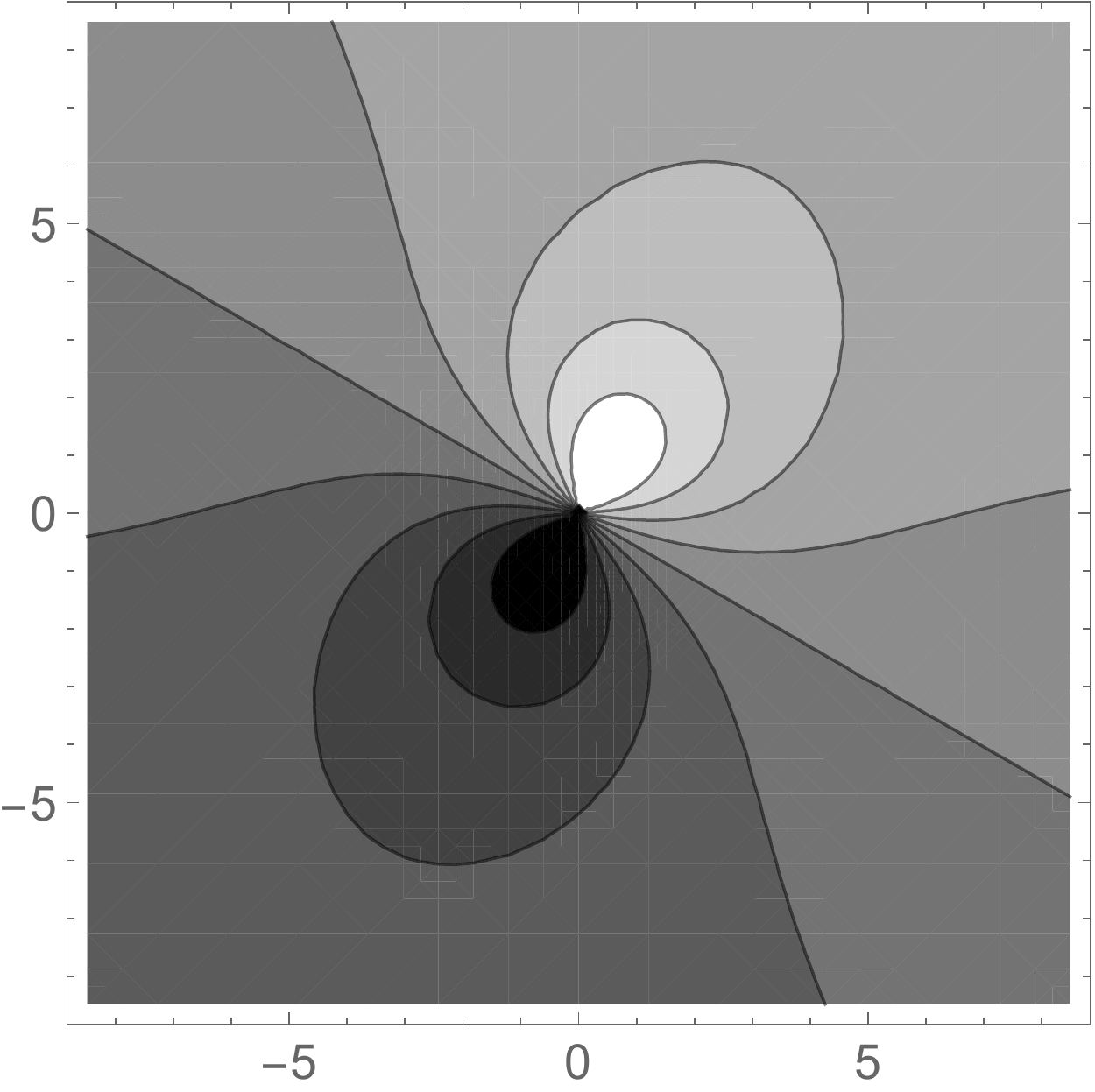}
\caption{$\chi^r$ at $kz = \pi/3$.}
\end{subfigure}%
\quad\quad\quad
\begin{subfigure}{5cm}
\centering
\includegraphics[width=0.3\linewidth]{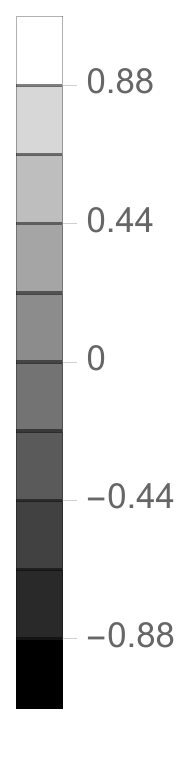}
\centering
\end{subfigure}
\caption{The graphs in a) b) and c) show the $\chi^r$ profiles as contour plots at $kz=0$, $\pi/6$, and $\pi/3$ respectively for $\eta = 2.2$.  The three plots indicate that the $\chi^r$ profile twists with pitch $\eta/2$.  This is also true of the other components $\chi^\theta$ and $\chi^z$.}
\end{figure}

\begin{figure}
\hspace*{1.3cm}
\centering
\begin{subfigure}{5cm}
\centering
\includegraphics[width=1.5\linewidth]{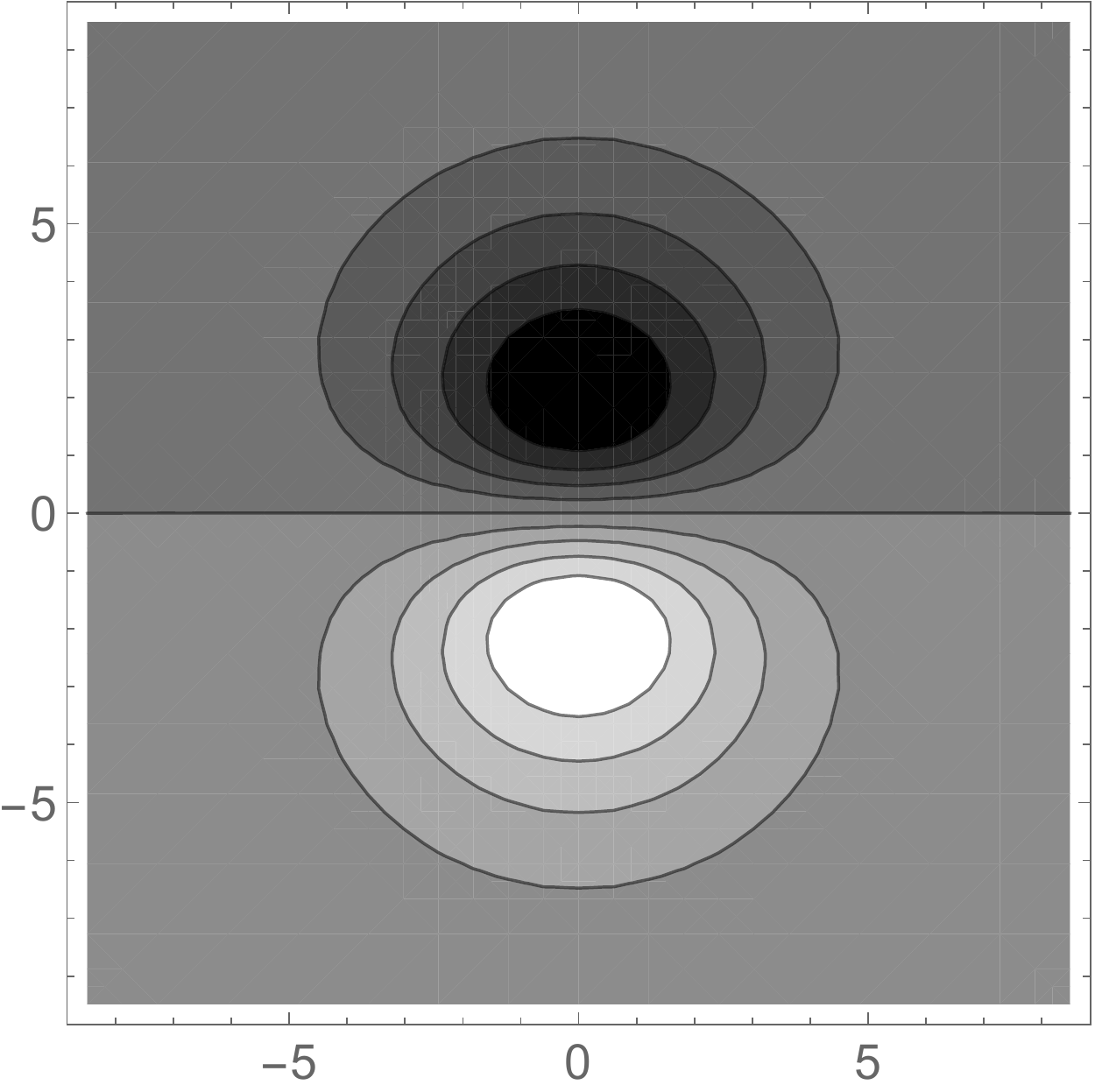}
\end{subfigure}%
\quad\quad\quad
\begin{subfigure}{5cm}
\centering
\includegraphics[width=0.35\linewidth]{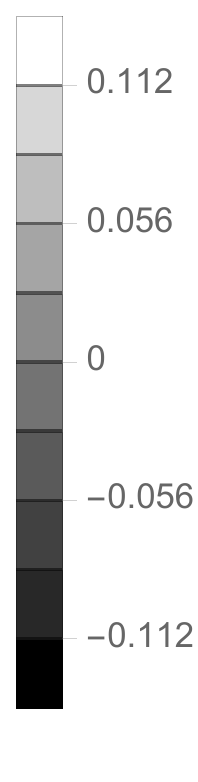}
\end{subfigure}%
\caption{Shown in this figure is a contour plot of $\chi^z$ at $kz = 0$, for $\eta=2.2$.  The plot indicates a dipole in the $\chi^z$ profile.}
\end{figure}

\begin{figure}
\hspace*{1.3cm}
\centering
\begin{subfigure}{5cm}
\centering
\includegraphics[width=1.5\linewidth]{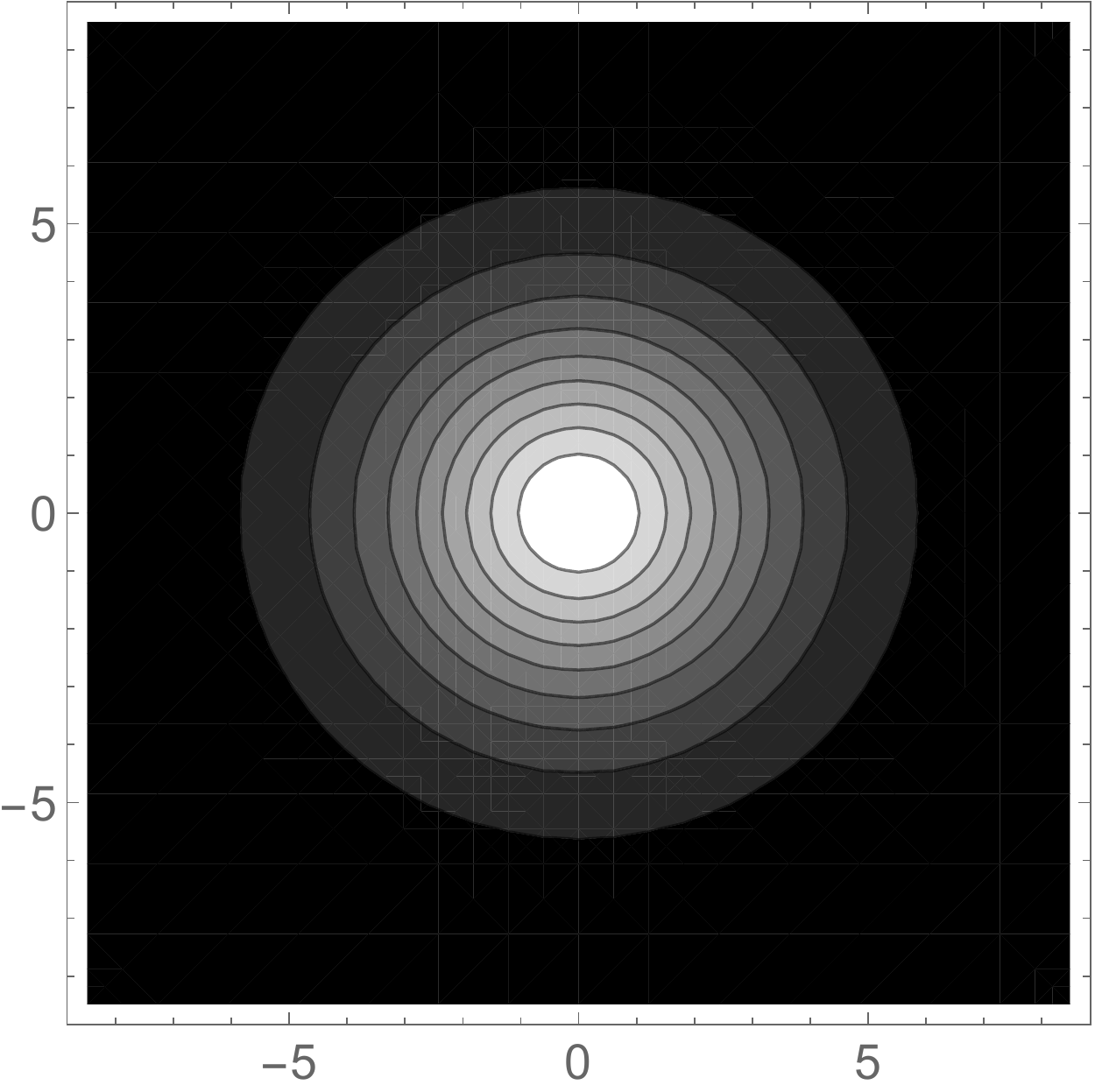}
\end{subfigure}%
\quad\quad\quad
\begin{subfigure}{5cm}
\centering
\includegraphics[width=0.27\linewidth]{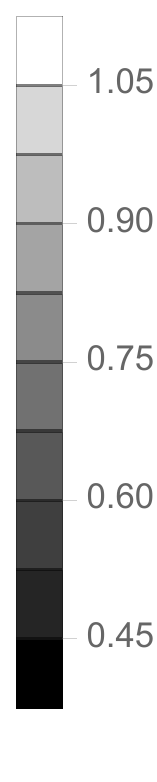}
\end{subfigure}%
\caption{Shown in this figure is a contour plot of $\chi = \sqrt{\chi^{r2}+\rho^2\chi^{\theta 2}+\chi^{z2}}$ at $kz = 0$, for $\eta=2.2$.  The plot indicates the near cylindrical symmetry of the field $\chi$.  This implies a minimal back reaction on $\phi$ and $\vec{A}$ due to the twisting of $\chi_i$.}
\end{figure}

\begin{figure}[h!]
\centering
\includegraphics[width=0.8\linewidth]{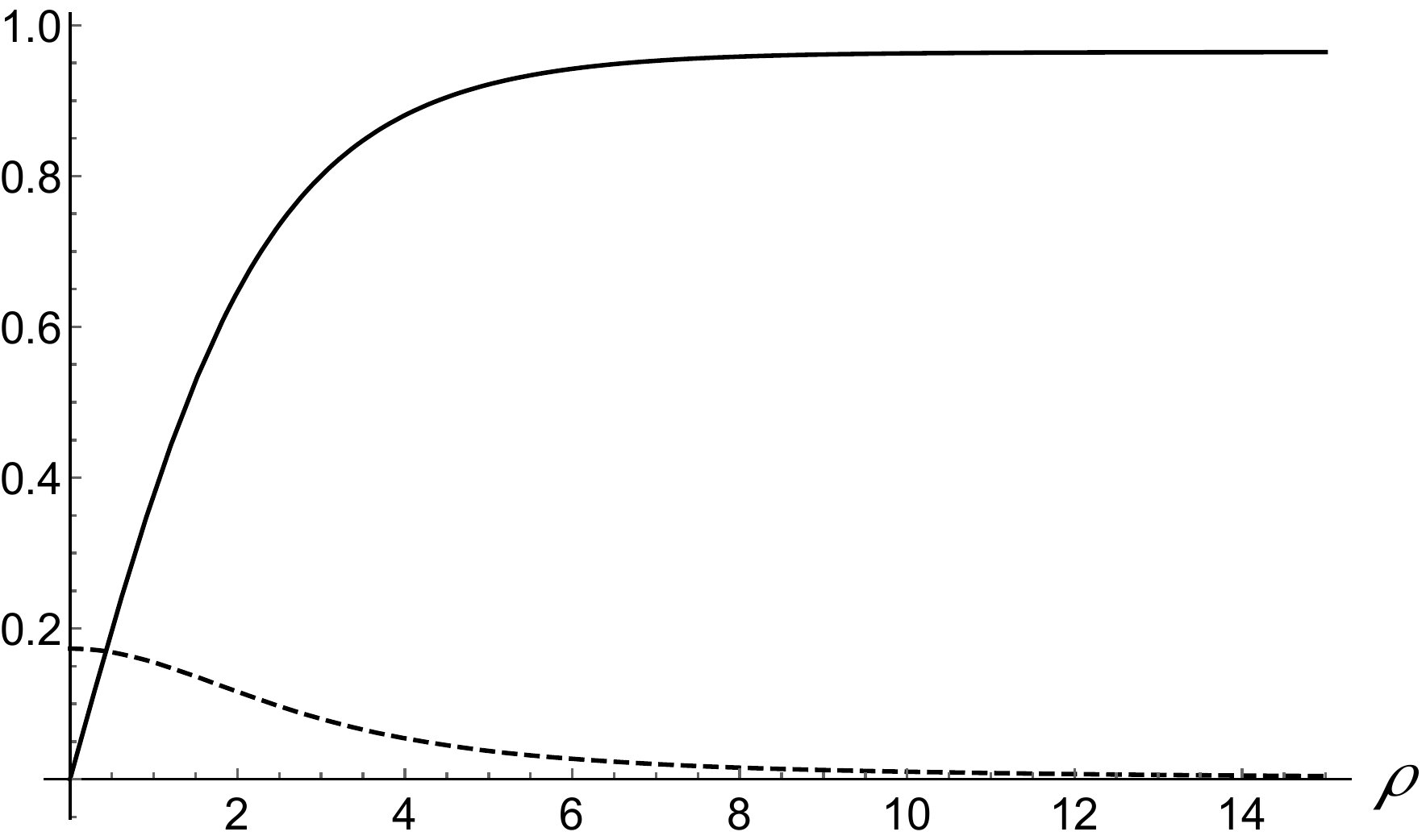}
\caption{$|\varphi(\rho)|$ (solid) and $A^\theta(\rho)$ (dashed) are shown at $z=\theta=0$ for $\eta = 2.2$.  The non-analytic behavior of $|\varphi|$ at $\rho \rightarrow 0$ indicates the $U(1)$ topological charge of the vortex solution.  The functions $A^r(\rho)$ and $A^z(\rho)$ are negligible in comparison to $A^\theta(\rho)$, and are thus not shown.}
\end{figure}

\begin{figure}
\hspace*{-2.4cm}
\begin{subfigure}{5cm}
\centering
\includegraphics[width=1.5\linewidth]{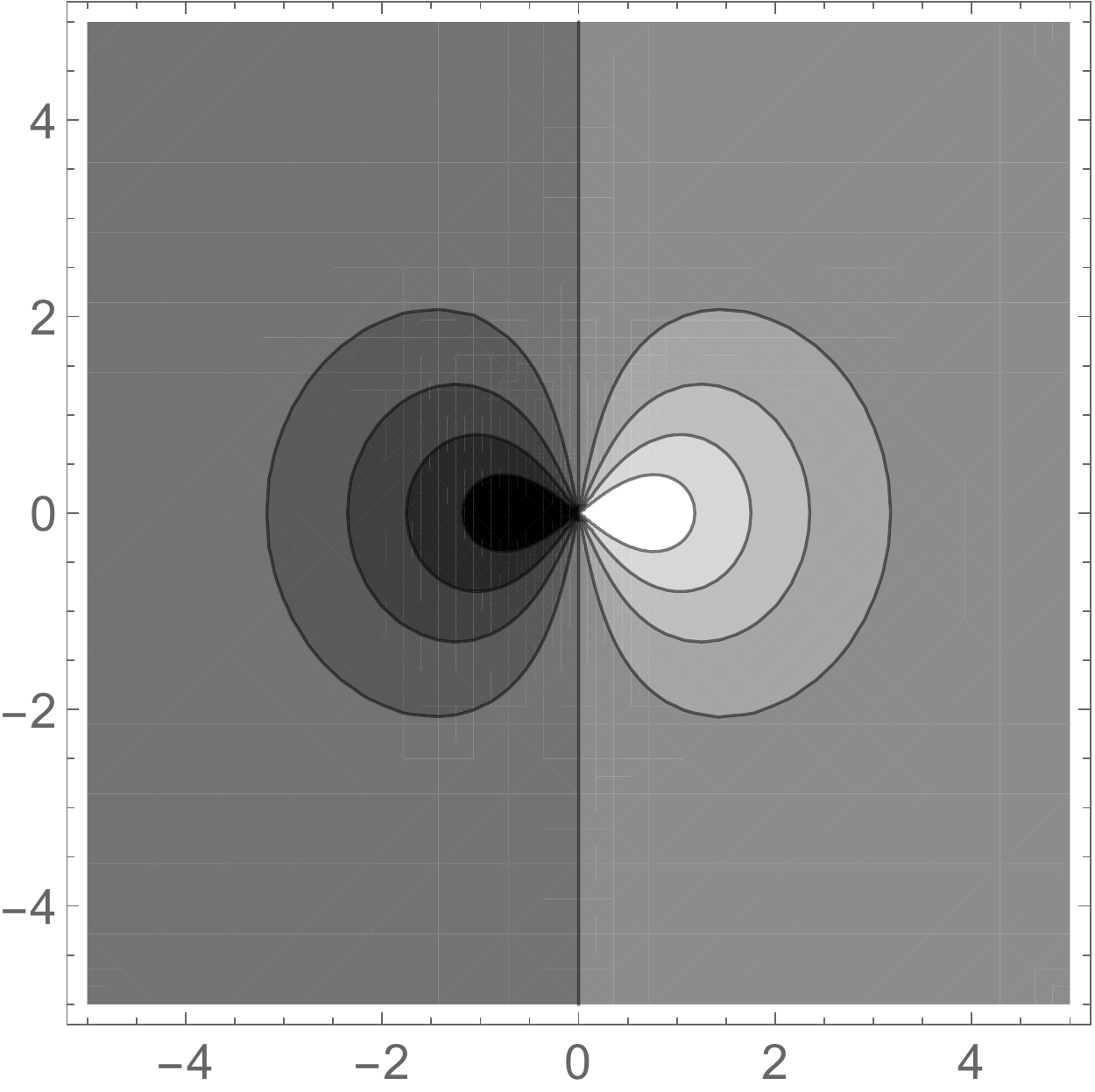}
\caption{$\chi^r$, at $kz=0$.}
\end{subfigure}%
\quad\quad\quad\quad\quad\;\;\;\;\;\;\;\;\;
\begin{subfigure}{5cm}
\centering
\includegraphics[width=1.5\linewidth]{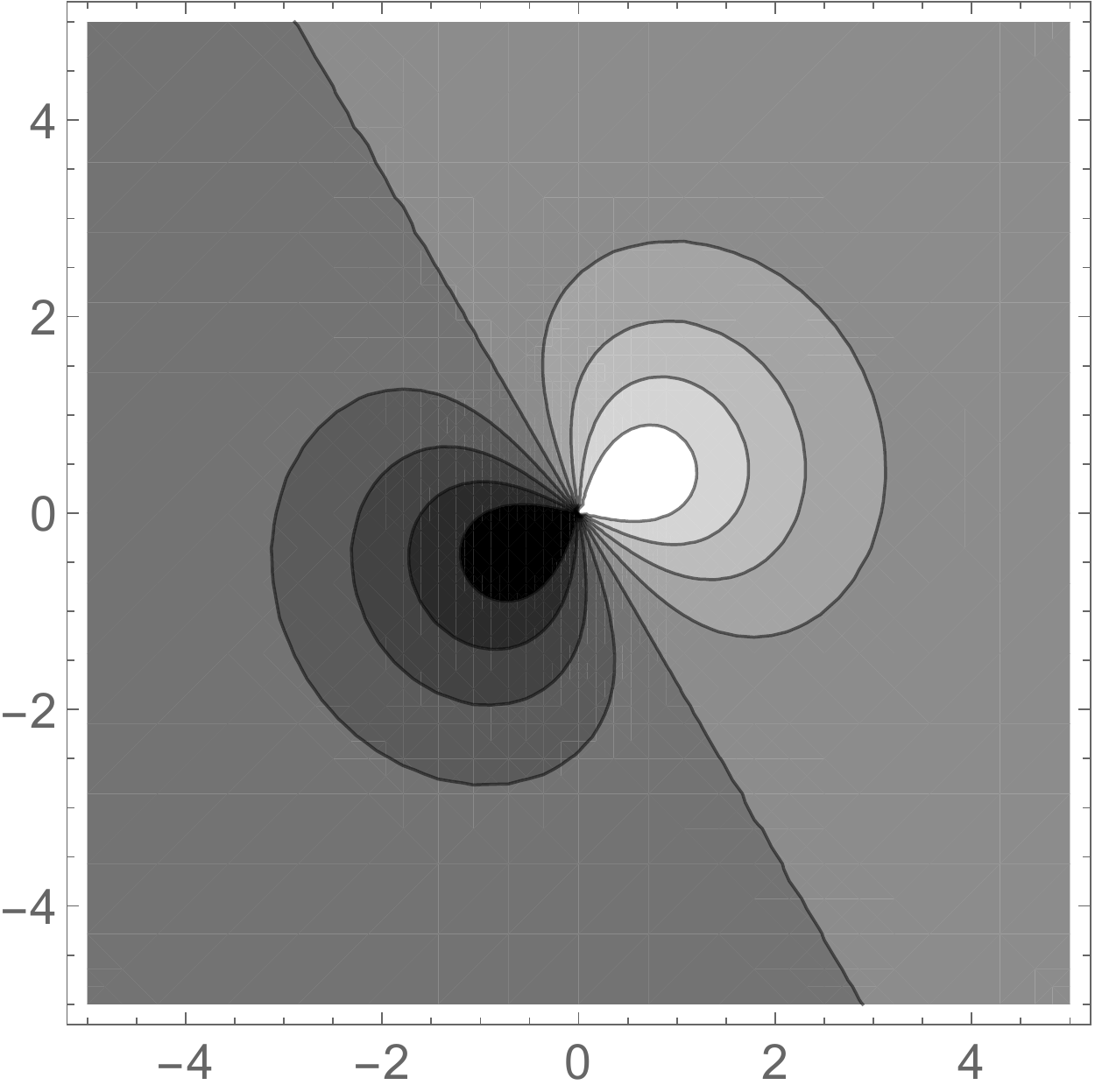}
\centering
\caption{$\chi^r$ at $kz = \pi/6$.}
\end{subfigure}\\
\\
\\
\centering
\hspace*{1.3cm}
\begin{subfigure}{5cm}
  \centering
\includegraphics[width=1.5\linewidth]{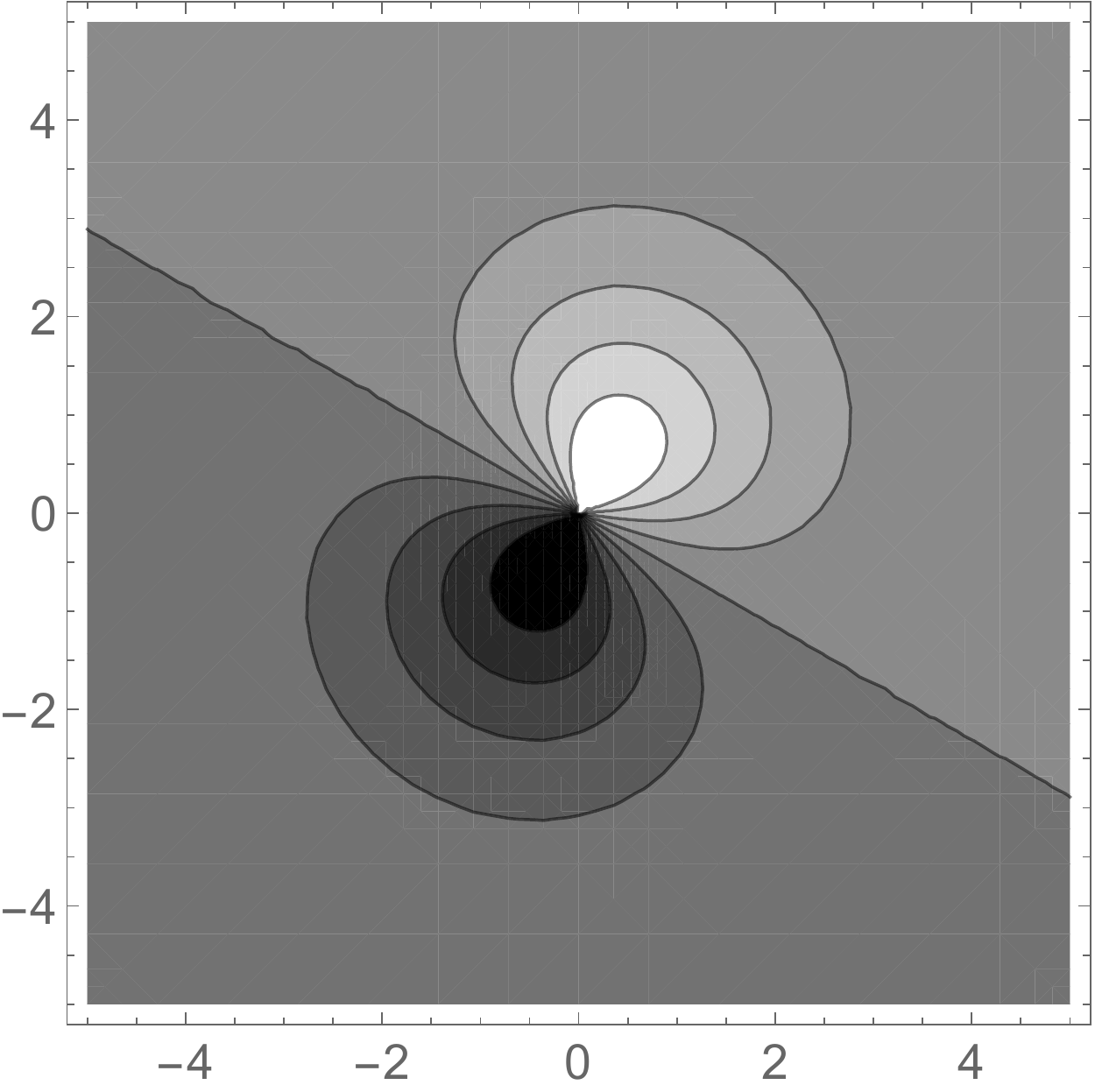}
\caption{$\chi^r$ at $kz = \pi/3$.}
\end{subfigure}%
\quad\quad\quad
\begin{subfigure}{5cm}
\centering
\includegraphics[width=0.3\linewidth]{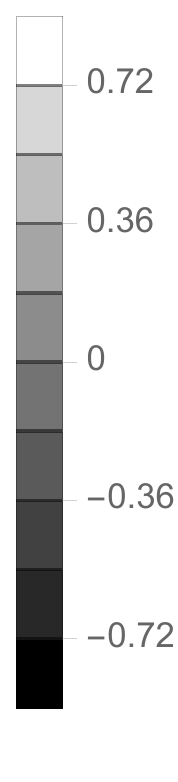}
\centering
\end{subfigure}
\caption{The graphs in a) b) and c) show the $\chi^r$ profiles as contour plots at $kz=0$, $\pi/6$, and $\pi/3$ respectively for $\eta = 0.5$.  Again the profile twists with pitch $\eta/2$.}
\end{figure}

\begin{figure}
\hspace*{1.3cm}
\centering
\begin{subfigure}{5cm}
\centering
\includegraphics[width=1.5\linewidth]{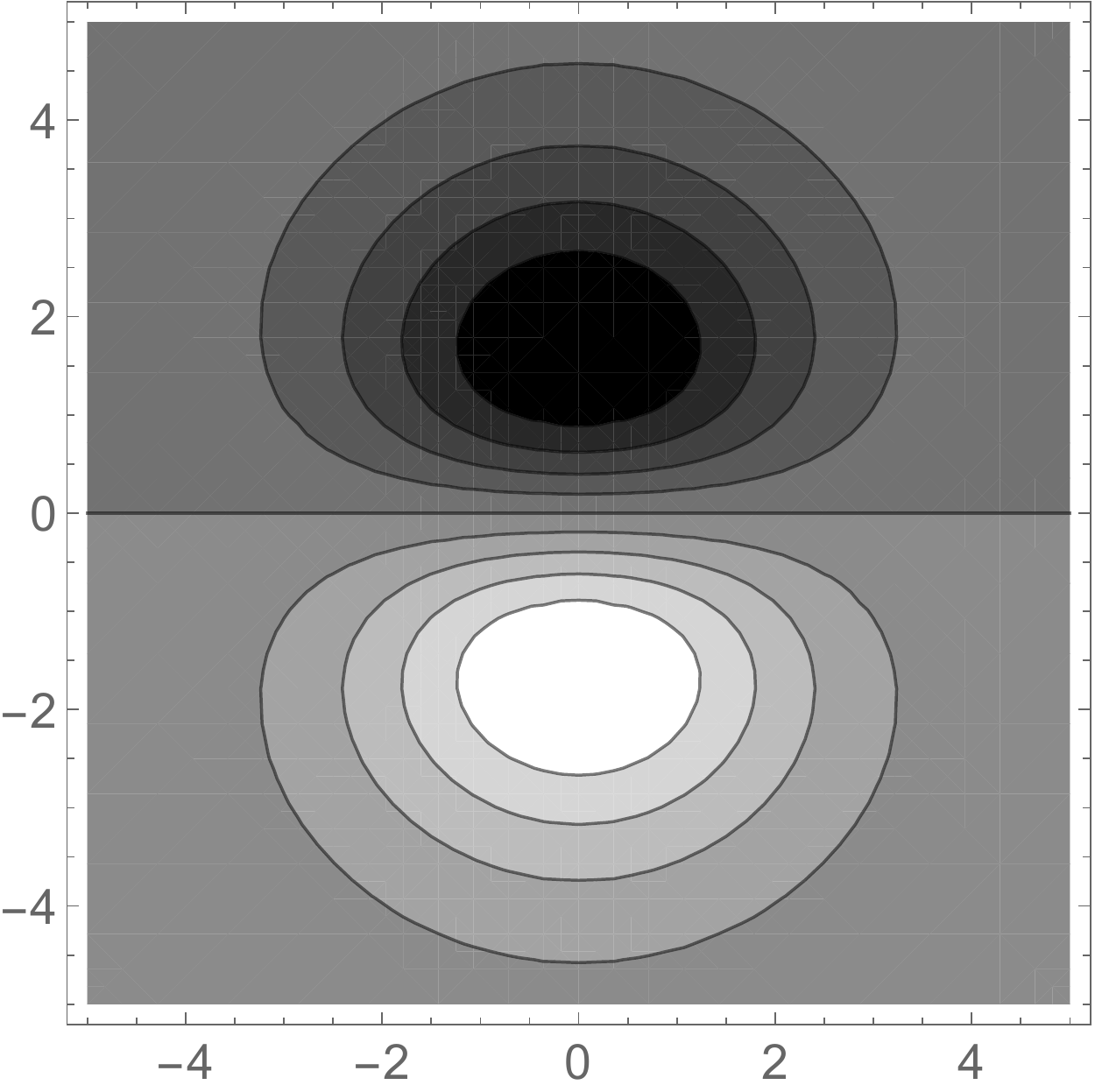}
\end{subfigure}%
\quad\quad\quad
\begin{subfigure}{5cm}
\centering
\includegraphics[width=0.35\linewidth]{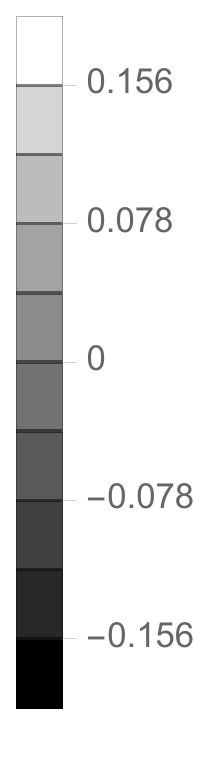}
\end{subfigure}%
\caption{Shown in this figure is a contour plot of $\chi^z$ at $kz = 0$, for $\eta=0.5$.  The dipole is again apparent in the $\chi^z$ profile.}
\end{figure}

\begin{figure}
\hspace*{1.3cm}
\centering
\begin{subfigure}{5cm}
\centering
\includegraphics[width=1.5\linewidth]{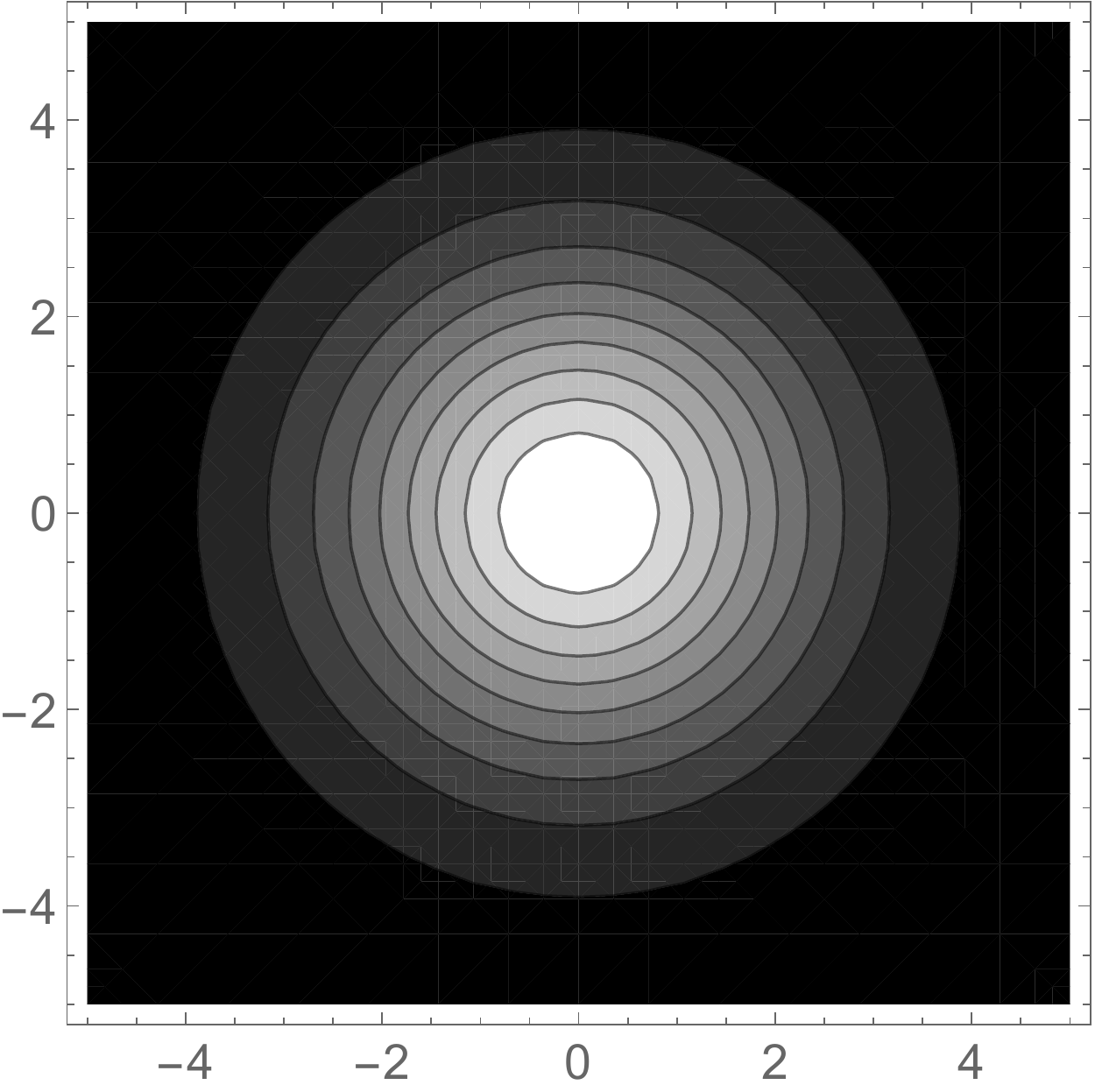}
\end{subfigure}%
\quad\quad\quad
\begin{subfigure}{5cm}
\centering
\includegraphics[width=0.3\linewidth]{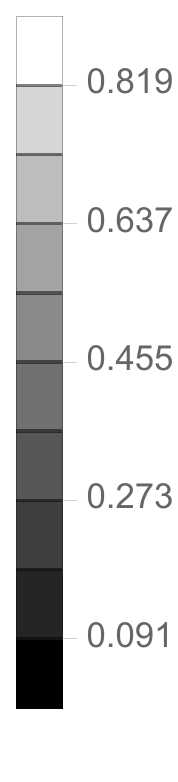}
\end{subfigure}%
\caption{Shown in this figure is a contour plot of $\chi = \sqrt{\chi^{r2}+\rho^2\chi^{\theta 2}+\chi^{z2}}$ at $kz = 0$, for $\eta=0.5$.  The plot indicates the near cylindrical symmetry of the field $\chi$.  This implies a minimal back reaction on $\phi$ and $\vec{A}$ due to the twisting of $\chi_i$.}
\end{figure}

\begin{figure}[h!]
\centering
\includegraphics[width=0.8\linewidth]{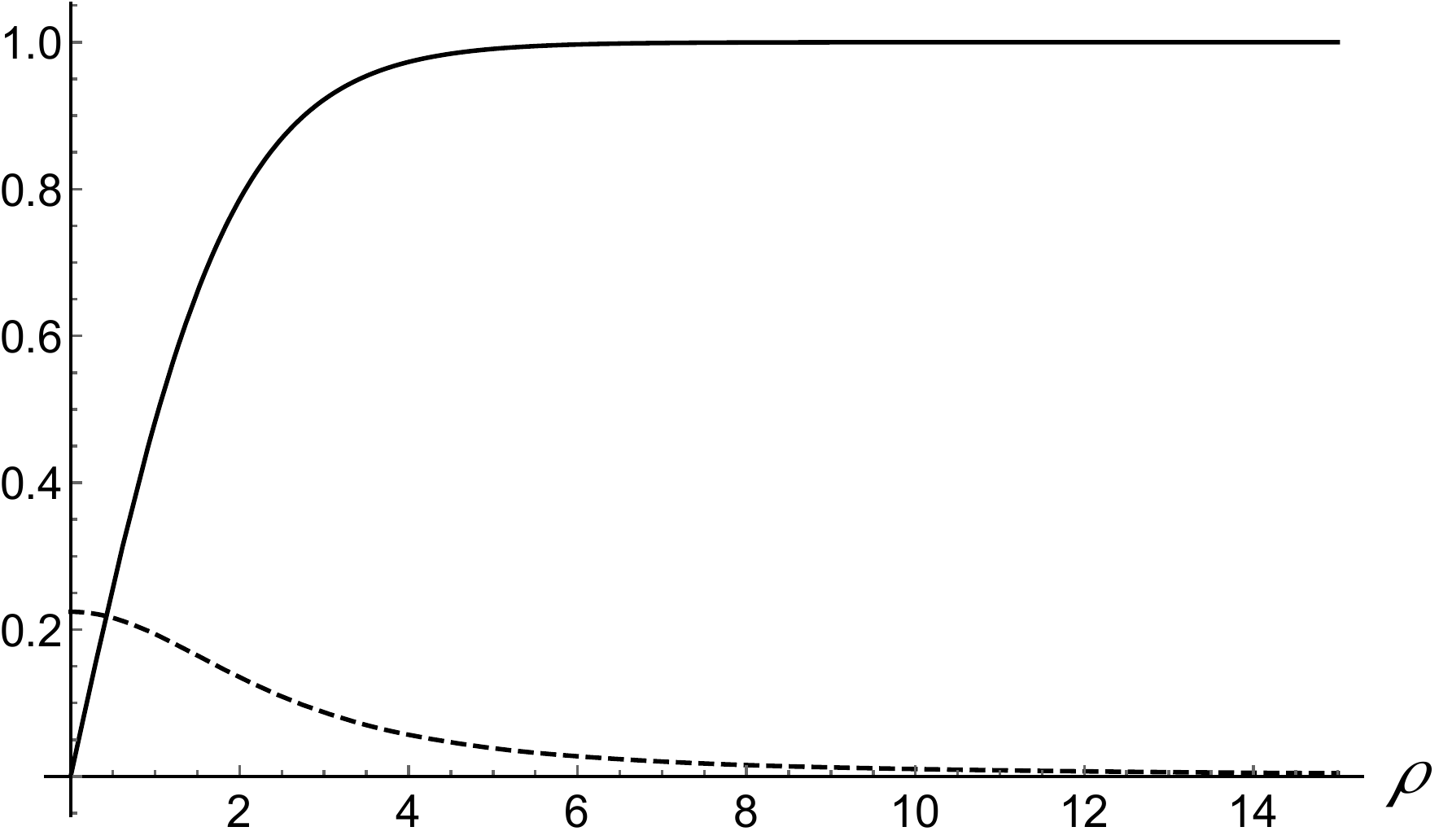}
\caption{$|\varphi(\rho)|$ (solid) and $A^\theta(\rho)$ (dashed) are shown at $z=\theta=0$ for $\eta = 0.5$.  Again, the functions $A^r(\rho)$ and $A^z(\rho)$ are negligible in comparison to $A^\theta(\rho)$.}
\end{figure}

\subsection{Numerical Technique}

The differential system (\ref{ChiSystem}), (\ref{ASystem}) and (\ref{PhiSystem}) for the dynamical vortex is a coupled set of eight differential equations that must be solved in three dimensions with the boundary conditions (\ref{DynamicalBoundaryConditions}).  In order to achieve an accurate numerical solution in a reasonable amount of computing time and memory, we make use of the pseudo-spectral method on a collocated grid \cite{Boyd:2001}.  We expand the solution in $\theta$ and $z$ coordinates in Fourier modes to exploit the periodicity expected in both coordinates.  For the radial coordinate we first compactify the coordinate so that the boundary condition at infinite radius can be implemented.  For this we transform the radial coordinate $r$ to the coordinate $x$ by
\begin{equation}
r = \frac{Lx}{\sqrt{1-x^2}}, 
\end{equation}
where $L$ is a constant we select by choice.  Typically we select $L= 2 \mbox{ or } 3$ to achieve the best distribution of grid points.  The radial solution is then expanded in Chebyshev polynomials $T_n(x)$ over a Radau collocation grid in $x$.  We implement our boundary condition at $x \rightarrow 1$, which corresponds to the large radius $r \rightarrow \infty$.  It is worth mentioning that the functions $T_n(x(r))$ are known as the rational Chebyshev functions.

Having written the solutions in terms of the expansion modes, we transform the differential equations into a coupled system of algebraic equations, which must be solved to determine the coefficients of the expansion modes.  To solve the large system, we use a Newton's method to achieve a solution iteratively until convergence is observed.  An added bonus of this method is that the stability properties of the approximate solution are shared with the linear stability of the exact solution for a large enough set of expansion modes \cite{Galantai:2000}.  Thus, if our method shows convergence to a stable approximation, we can be sure of the linear stability of our solution to small perturbations.

\section{Low energy theory of vortex moduli}

It is well known that in the absence of a cholesteric structure, when $\eta=0$, the system with energy density (\ref{EnergyDensityEta}) describes non-Abelian strings. The low energy theory on their world-sheet is a $CP(1)$ model for two orientational moduli (plus the standard translational moduli of the vortex centre). When $\eta \neq0$, once one aligns the vortex axis with the cholesteric wave-vector, the $O(3)$ rotational freedom of $\chi_i$ is broken to the subgroup of rotations in the plane orthogonal to $\vec{k}$. Correspondingly some of the orientational moduli have to be lifted.

To determine the low energy theory at the classical level, we rely on the symmetries of the system following the procedure in \cite{Nitta:2013}.  In particular, the moduli appearing on the vortex are determined by the continuous global symmetries of the vacuum, which are broken by the vortex core.  The continuous global symmetry group of the Lagrangian (\ref{Lagrangian}) with the parity breaking term (\ref{LagrangianEta}) is given by
\bea
G_{\rm global} = SO(3)_{J} \times T,
\eea 
where $T$ is the three dimensional translation group, and $J$ refers to the sum of generators $S+L$.  

The vacua appearing in the model preserve part or all of the symmetry group $G$.  For vacuum $I$ in (\ref{Vac1}) the appearance of a non-zero $\chi_0$ implies the symmetry breaking of translations in the direction of the wave vector $\vec{k} \equiv k\hat{z}$.  Thus only the subgroup $T_x \times T_y$ of the translational group $T$ remains invariant in this vacuum.  Additionally, since $\vec{\chi}_0$ rotates in the $xy$-plane as one translates in the $z$ direction, rotations about the $z$ direction are broken.  The wave vector $\vec{k}$ breaks the remainder of the $SO(3)_{J}$ group.  However, a glance at (\ref{CholestericStructure}) shows that translations in the $z$ direction are equivalent to rotations about $z$ in vacuum $I$.  Thus, although vacuum $I$ separately breaks $T_z$ and $SO(2)_{J_z}$, a subgroup, which we will denote $SO(2)_{J'_z} \subset SO(2)_{J_z} \times T_z$ is preserved.  Thus the complete symmetry of vacuum $I$ is given by
\bea
H_{I} = T_{x,y} \times SO(2)_{J'_z}.
\eea

Vacuum $II$ on the other hand does not break the translational or rotational symmetries since $\chi_0 = 0$ in this case.  Thus,
\be
H_{II} = G_{\rm global}.
\ee

The continuos global symmetries preserved by vacuum $III$ are exactly the same as those of $I$.  Vacuum $III$ is only distinguished from $I$ because $III$ does not break the $U(1)$ gauge symmetry.
\bea
H_{III} = H_{I}.
\eea

The presence of vortices in these vacua further break the symmetry of the system.  The generators of the vacuum subgroups $H_i$ which are broken by the vortex solution produce the moduli fields in the low energy theory.  With this in mind we may write down a low energy theory by varying the vortex solutions through their degeneracy space.  For the translations in the $x$ and $y$ directions this can be achieved by replacing in the Lagrangian
\begin{align}
&\chi_i(\vec{x})\rightarrow \chi_i(\vec{x}-\vec{\xi}(t,z)), \nonumber \\
&\varphi(\vec{x}) \rightarrow \varphi(\vec{x}-\vec{\xi}(t,z)), \nonumber \\
&A_i(\vec{x}) \rightarrow A_i(\vec{x}-\vec{\xi}(t,z)),
\label{Translations}
\end{align}
and expanding to second order in $\vec{\xi}(t,z)$.  After integrating over $x$ and $y$ one achieves an effective Lagrangian in the modulus fields $\vec{\xi}(t,z)$.  If $\eta = 0$ one may also expand the vortex solution in the orientational modulus $\vec{S}(t,z)$ where the field $\chi_i$ is replaced in the Lagrangian
\be
\chi_i \rightarrow R_{ij}(t,z) \chi_j \equiv S_i(t,z)\chi(\vec{x}_\perp),
\ee
where $\vec{x}_\perp \equiv (x,y)$ are the coordinates perpendicular to the vortex axis.  $R_{ij}$ is the spin rotation matrix in $SO(3)_S$.  We also define $\chi^2 \equiv |\vec{\chi}|^2$.  Thus, $\vec{S}(t,z)$ satisfies the constraint $|\vec{S}|^2=1$.  Thus, when $\eta = 0$ the effective theory is given by
\be
S_{\eta = 0} = \int dtdz \left\{\ \frac{1}{2g^2}\partial_\alpha S_i \partial_\alpha S_i+\frac{T}{2} \partial_\alpha \xi_b\partial_\alpha \xi_b \right\}\,
\ee
where $g$ and $T$ are calculated from the integration over $x$ and $y$ \cite{Shifman:2013a}.  Here $\alpha = (t,z)$, and $b = (x,y)$.  The component $\xi_z$ does not appear in the low energy theory since the vortex solution in this case is independent of $z$.

We now switch on $\eta \neq 0$ and observe the effects on the vortex moduli in the various vacua.  For this we focus our attention on the $\chi_i$ kinetic terms in the Lagrangian.  For convenience in determining the moduli fields we write 
\be
\chi_i \equiv \chi(t,\vec{x}) S_i(t,\vec{x}), \;\; \mbox{ where } |\vec{S}|^2 \equiv 1.
\ee
This definition allows for a convenient form of the kinetic terms in the Lagrangian,
\be
\mathcal{L}_{\rm kin} = \partial^\mu \chi \partial_\mu \chi +\chi^2\left(\partial^\mu S_i\partial_\mu S_i -\eta\varepsilon _{ijk}S_i \partial_j S_k\right).
\label{KineticLagrangian}
\ee

Let us first consider the vortices in vacuum $I$.  Obviously, the translational moduli $\xi_{x,y}(t,z)$ are still existent.  However, since the solutions no longer preserve $z$ translational symmetry, the additional translations $\xi_z(t,z)$ along the $z$-direction along the vortex axis may appear.  However, the vacuum far from the vortex axis is characterized by a non-zero $\chi_i \rightarrow \chi_0 \epsilon_i(z)$.  Thus, the kinetic term for $\xi_z(t,z)$ in the action is given by:
\be
S_{\xi_z} = \frac{T_z}{2} \int dtdz (\partial_\alpha \xi_z)^2
\ee
where the tension $T$ is given by
\be 
T_z \rightarrow \int d^2\vec{x}_\perp \chi_0^2,
\ee
which is obviously divergent.  This is expected since the vacuum $I$ only preserves a translational symmetry in the $z$ direction if it is accompanied by a corresponding rotation in $SO(2)_{J_z}$.  However, for the parameters considered in the previous sections the vortex solutions also satisfy the $SO(2)_{J'_z}$  symmetry to a very good approximation and thus cannot produce a dynamical modulus from this group.  For larger back reactions (due to large $\eta >> \eta_{{\rm crit}_2}$ for example) on the $\phi$ and $A_i$ fields this approximation will no longer hold, and an $SO(2)_{J'_z}$ modulus will appear in the low energy dynamics.  This $SO(2)_{J'_z}$ modulus can be shown to emerge on the $\phi$ field due to the back reaction from the last term proportional to $\chi^2$ in $V_\varphi$ in (\ref{Potentials}).  For large $\eta$, $\chi_i$ will have a dipole term with a $\theta-kz$ dependence, and thus $\chi^2$ will have higher multipoles (at least quadrupole) appearing in a Fourier expansion in $\theta' \equiv \theta-kz$.  This will break the $SO(2)_{J'_z}$ symmetry, and lead to an additional modulus appearing on the $\phi$ vortex solution.  We will not consider this case in further detail here.

The more interesting case occurs in vacuum $II$ where the entire global symmetry group $G_{\rm global}$ is preserved in the vacuum, and only the $U(1)$ gauge symmetry is broken.  We will assume for this case that $0< \eta << \eta_{{\rm crit}_1}$.  With this constraint we may assume a vortex solution of the form
\be
\chi_i \approx \chi(\rho) \epsilon_i(z),
\ee
where the modulus $\chi(\rho)$ has near axial symmetry.  We now apply the translational transformations (\ref{Translations}) and the $SO(3)_J$ transformations,
\be
\epsilon_i \rightarrow R_{ij}(\vec{\omega}(t,z))\epsilon_j \equiv S_i(t,z).
\label{Rotations}
\ee
Expanding (\ref{KineticLagrangian}) to second order in $\xi(t,z)$ and integrating over $x$ and $y$ we find
\begin{align}
\mathcal{L}_{\rm kin} =\frac{T}{2}(\partial_\alpha \vec{\xi}_\perp)^2+ \frac{1}{2g^2}&\left[ (\partial_\alpha \vec{S})^2+ (\partial_\alpha \vec{S})^2(\partial_z \xi_z)^2+\eta\varepsilon_{ab}S_a\partial_zS_b  \right. \nonumber \\
&\left. +2\partial_z \xi_z \left((\partial_z \vec{S})^2 -\frac{\eta}{2}\varepsilon_{ab}S_a \partial_z S_b\right) \right],
\label{LowEnergyTheory}
\end{align}
where $\alpha = (t,z)$, and $(a,b)$ are indices in $(x,y)$.  The constant $g$ is calculated from the integration over $x$ and $y$,
\be
\frac{1}{2g^2} = \int d^2\vec{x}_\perp \chi(\rho)^2.
\ee
We emphasize for clarity that $\vec{S}$ in (\ref{LowEnergyTheory}) has both the non-perturbative dynamical structure $\epsilon_i(z)$ from (\ref{CholestericStructure}) in addition to the small perturbations on the ground state.

We first note that the second line of (\ref{LowEnergyTheory}) can be expanded to first order in $\vec{\omega}$ with the result
\be
2\partial_z \xi_z \left((\partial_z \vec{S})^2 -\frac{\eta}{2}\varepsilon_{ab}S_a \partial_z S_b\right)=-\eta\partial_z \xi_z \partial_z \omega_z+\mathcal{O}(\vec{\omega}^2).
\ee
Combining this with the $\mathcal{O}(\xi_z^2)$ and $\mathcal{O}(\omega_z^2)$ in the first line of (\ref{LowEnergyTheory}) we find
\be
\mathcal{L}_{\rm kin} \supset \frac{1}{2g^2}\left[\partial_z\left(\frac{\eta}{2}\xi_z + \omega_z\right)\right]^2 \equiv \frac{1}{2g^2}\left(\partial_z\omega'_z\right)^2,
\label{CombinedModulus}
\ee
which simply enforces the (approximate) $SO(2)_{J'_z}$ symmetry of the vortex.  Thus, only one of the two moduli coming from the broken $z$ translations and the $SO(2)_{J_z}$ rotations survives as a low energy dynamical field.

The last observation we wish to discuss is the remaining two rotational moduli $\omega_{x,y}$ from the degeneracy space $SO(3)_{J_z}/SO(2)_{J_z}$.  Expanding the first line of (\ref{LowEnergyTheory}) to second order in $\omega_{x,y}$ we arrive at
\be
\partial_zS_i\partial_zS_i =\left((\partial_z\omega_x+k\omega_y)\sin kz - (\partial_z\omega_y - k \omega_x)\cos kz\right)^2 +\mathcal{O}(\omega^3).
\label{OmegaKineticTerms}
\ee
To obtain this formula we expand $S_i$ to first order in $\omega$ using
\be
\vec{S} = {\rm e}^{{\rm i}\vec{\omega}\cdot\vec{L}}\vec{\epsilon} \approx \left(1 + {\rm i}\vec{\omega}\cdot\vec{L}\right)\vec{\epsilon} =\left(\begin{array}{c} 
\cos kz-\omega_z  \sin kz\\
\sin kz+\omega_z \cos kz  \\
\omega_x \sin kz-\omega_y \cos kz \\
\end{array}\right).
\ee
Taking the derivative of this expression and squaring gives us the formula (\ref{OmegaKineticTerms}).  

We observe for $\partial_z \omega_a \gg \eta \omega_a$ that
\be
(\partial_z S_a)^2 \approx (\partial_z\omega_y)^2.
\label{HighE}
\ee
Thus, for high energy excitations in $\omega_a$, the modulus $\omega_y$ survives as a massless degree of freedom.  

On the other hand if $\partial_z\omega_i \ll \eta\omega_a$ we have
\be
(\partial_zS_i)^2 \approx \frac{\eta^2}{4}\left(\omega_x^2+\omega_y^2\right)+(\mbox{derivatives of } \omega_a).
\label{LowE}
\ee
As expected, at low energies the two moduli from $SO(3)_{J_z}/SO(2)_{J_z}$ are lifted due to the emerging mass terms in $\omega_{x,y}$.

The interpretation of the high energy result (\ref{HighE}) is that at high momenta, the twisting structure of the vortex is decoupled from the excitations, and the original sigma model for $\eta \rightarrow 0$ reappears approximately.  However, low energy excitations see an average over many cycles of the vortex twisting leading to a lifting of the moduli $\omega_{x,y}$ (see \cite{Radzihovsky:2011}).  We point out that the high energy result (\ref{HighE}) cannot be disregarded on the grounds of classical low energy dynamics.  Indeed it is quite possible for the energy of the excitation to be larger than $\eta$ while still being low energy relative to the width of the vortex.

\section{Discussion and Conclusions}

We have observed some of the effects of a parity violating twist term in the Lagrangian (\ref{LagrangianEta}), which is linear in the derivative.  The most important of these observations is a new cholesteric vacuum field configuration which break translational and rotational symmetry.  In addition, we have discussed vortex solutions with the gauge $U(1)$ topological charge in the two vacua ($I$ and $II$) where such vortices are allowed classically.  

We have deduced the essentials of the classical low energy excitations of such vortices.  This was achieved by appealing to symmetry arguments when the twist term can be treated as a perturbation of the original Lagrangian (\ref{Lagrangian}).  In particular, we have observed the emergence of a new translational modulus $\omega'_z$ appearing in (\ref{CombinedModulus}).  In addition, we have observed the lifting of the non-Abelian moduli $\omega_{x,y}$ of the original $(1+1)$ dimensional $O(3)$ sigma model.  For the case of vortices in vacuum $II$ this occurs for low energy excitations of $\omega_{x,y}$ which develop a mass proportional to $\eta$.  On the other hand, excitations with momenta much larger than $\eta$ probe the vortex line at small enough distances where the twist effect decouples.

This paper has focused on the dynamical effects on vortices with a $U(1)$ topological charge after introducing a parity violating twist term in the Lagrangian (\ref{Lagrangian}).  However, several topics of this system remain unanswered and could provide avenues of future research.  Specifically, although we have mentioned the existence of vortices with a non-Abelian global $SO(3)_{S+L}$ charge, we have not discussed their solutions in detail.  These solutions are the so called disclinations.  They are similar to the vortices in a biaxial nematic liquid crystal where the non-zero value of $\eta$ and $\chi_0$ leads to a degeneracy space $SO(3)$ with non-trivial fundamental group.  The details of these vortices in cholesteric liquid crystals have been worked out (see \cite{deGennes:1993} or \cite{Mineev:1998} for example), however their properties in the present context, with the additional $U(1)$ gauge symmetry, are yet to be discussed (although see \cite{Kondo:1992}).

Our discussion of the low energy excitations has been confined to a classical discussion.  We have not considered the quantization of the zero modes, which is a non-trivial topic when the system is not Lorentz invariant \cite{Watanabe:2012} and spacetime symmetries are broken \cite{Low:2001}.

\section*{Acknowledgements}
The work of A.P. is supported by the Doctoral Dissertation Fellowship at the University of Minnesota.  The work of M.S. is supported in part by DOE Grant Number DE-SC0011842.  G.T. is funded by Fondecyt Grant Number 3140122.  The Centro de Estudios Cient\'{i}ficos (CECS) is funded by the Chilean Government through the Centers of Excellence Base Financing Program of Conicyt.  G.T. would like to thank the Fine Institute for Theoretical Physics at the University of Minnesota for hospitality during the completion of this work.

\end{document}